\documentclass[12pt]{article}
\pdfoutput=1
\usepackage{jheppub, amsmath,amssymb,amsfonts,amsxtra, mathrsfs, makeidx,graphics,graphicx,amsthm,epsfig, ytableau,bm,longtable,float, color,array, xkeyval,adjustbox,collectbox,ifpdf, tikz, mathtools, paralist}
\usetikzlibrary{positioning}
\usetikzlibrary{chains}

\usepackage{bbm}

\newcommand{\ie}{\textit{i.e.}}

\numberwithin{equation}{section}

\newcommand{\nn}{\nonumber}

\newcommand{\be}{\begin{equation}} \newcommand{\ee}{\end{equation}}
\newcommand{\bea}{\begin{equation} \begin{aligned}} \newcommand{\eea}{\end{aligned} \end{equation}}
\newcommand{\bsp}{\begin{split}} \newcommand{\esp}{\end{split}}

\def\tilde{\widetilde}

\def\rt2{\sqrt{2}}

\def\det{\mathop{\rm det}}

\def\Tr{\mathop{\rm Tr}}

\def\CH{{\cal H}}

\def\CM{{\cal M}}
\def\CN{{\cal N}}

\def\CZ{{\cal Z}}

\def\1{{\ds 1}}

\newcommand{\cG}{\mathcal{G}}

\newcommand{\cN}{\mathcal{N}}

\newcommand{\cR}{\mathcal{R}}

\newcommand{\cW}{\mathcal{W}}

\newcommand{\cZ}{\mathcal{Z}}

\newcommand{\bC}{\mathbb{C}}

\newcommand{\bR}{\mathbb{R}}
\newcommand{\bZ}{\mathbb{Z}}

\newcommand{\unit}{\mathbbm{1}}

\newcommand{\tO}{{\tilde O}}

\def\repa{\raise4pt\hbox{$\square$}\mkern-14mu\raise-4pt\hbox{$\square$}}
\def\repab{\overline{\raise4pt\hbox{$\square$}\mkern-14mu\raise-4pt\hbox{$\square$}\mkern-1mu}}

\def\smileface{\ensuremath{\hbox{\large$\bigcirc$}\mkern-15mu\raise-1pt\hbox{\scriptsize$\smallsmile$}%
\mkern-10mu\raise4pt\hbox{..}\mkern4mu}}
\def\frownface{\ensuremath{\hbox{\large$\bigcirc$}\mkern-15mu\raise-1pt\hbox{\scriptsize$\smallfrown$}%
\mkern-10mu\raise4pt\hbox{..}\mkern4mu}}

\newcommand{\ba}{\begin{array}}
\newcommand{\ea}{\end{array}}
\newcommand{\bi}{\begin{itemize}}
\newcommand{\ei}{\end{itemize}}
\def\vec#1{\bm{#1}}
\def\bea#1\eea{\allowdisplaybreaks \begin{align}#1\end{align}}
 \newcommand{\ben}{\begin{enumerate}}
\newcommand{\een}{\end{enumerate}}
\newcommand{\bean}{\begin{eqnarray*}}
\newcommand{\eean}{\end{eqnarray*}}
\newcommand{\eref}[1]{(\ref{#1})}

\newcommand{\tref}[1]{Table~\ref{#1}}
\newcommand{\fref}[1]{Figure~\ref{#1}}
\newcommand{\PE}{\mathop{\rm PE}}
\newcommand{\PL}{\mathop{\rm PL}}

\newcommand{\BC}{\mathbb{C}}

\newcommand{\BZ}{\mathbb{Z}}

\newcommand{\comment}[1]{}
\newcommand{\diag}{\mathrm{diag}}

\newcommand{\adj}{\mathbf{Adj}}

\newcommand{\Adj}{\mathbf{\adj}}

\newcommand{\Sym}{\mathrm{Sym}}

\newcommand{\blue}{\color{blue}}

\def\node#1#2{\overset{#1}{\underset{#2}{\circ}}}
\def\Node#1#2{\overset{#1}{\underset{#2}{ \bullet}}}
\def\ver#1#2{\overset{{\llap{$\scriptstyle#1$}\displaystyle\circ{\rlap{$\scriptstyle#2$}}}}{\scriptstyle\vert}}

\usetikzlibrary{arrows,fit,decorations.pathreplacing}
\tikzstyle{every picture}+=[remember picture]
\tikzstyle{na} = [baseline=-.5ex]

\title{Coulomb Branch and The Moduli Space of Instantons}
\author[a]{Stefano Cremonesi,}
\author[a]{Giulia Ferlito,}
\author[a]{Amihay Hanany,}
\author[b]{and Noppadol Mekareeya}

\affiliation[a]{Theoretical Physics Group, Imperial College London, \\
Prince Consort Road, London, SW7 2AZ, UK}
\affiliation[b]{Theory Division, Physics Department, CERN, \\CH-1211, Geneva 23, Switzerland}

\emailAdd{s.cremonesi} 
\emailAdd{giulia.ferlito11}
\emailAdd{a.hanany@imperial.ac.uk}
\emailAdd{noppadol.mekareeya@cern.ch}

\preprint{
{\small
\begin{flushright}
Imperial/TP/14/AH/08\\
CERN-PH-TH-2014-136
\end{flushright}
}
}

\abstract{The moduli space of instantons on $\BC^2$ for any simple gauge group is studied using the Coulomb branch of $\CN=4$ gauge theories in three dimensions.  For a given simple group $G$, the Hilbert series of such an instanton moduli space is computed from the Coulomb branch of the quiver given by the over-extended Dynkin diagram of $G$.  The computation includes the cases of non-simply-laced gauge groups $G$, complementing the ADHM constructions which are not available for exceptional gauge groups.   Even though the Lagrangian description for non-simply laced Dynkin diagrams is not currently known, the prescription for computing the Coulomb branch Hilbert series of such diagrams is very simple. For instanton numbers one and two, the results are in agreement with previous works. New results and general features for the moduli spaces of three and higher instanton numbers are reported and discussed in detail.}

\begin{document}
\maketitle
\section{Introduction}

Instantons were first introduced as Euclidean finite action solutions of the self-dual pure Yang-Mills equations \cite{Belavin:1975fg,PhysRevD.14.3432}. 
The space of such solutions, graded by an integer number $k$, the Pontryagin number (or charge) of the instanton, is known as the moduli space of instantons. An algebraic prescription to construct instanton solutions for classical gauge groups $SU(N)$, $SO(N)$, $USp(2N)$ on $\bR^4$ was developed by Atiyah, Drinfeld, Hitchin and Manin in \cite{Atiyah:1978ri}. With the advent of $D$-branes as dynamical objects, the ADHM construction was given geometric light by means of a brane realization \cite{Witten:1995gx, Douglas:1995bn}: 
$Dp$-branes inside $D(p+4)$-branes are codimension $4$ objects, which dissolve into instantons for the worldvolume gauge fields of the $D(p+4)$-branes. 
For the gauge theory living on the $Dp$-brane, which has 8 supercharges, the Higgs branch of the moduli space therefore corresponds to the moduli space of instantons of the $D(p+4)$ gauge group.

In order to compute moduli spaces of instantons for classical gauge groups, one avenue is thus analyzing the Higgs branch of the ADHM quiver gauge theory. This is done by considering the constraints given by the $F$ and $D$ terms in the supersymmetric gauge theory and modding out by the gauge group. The Higgs branch for theories with 8 supercharges is classically exact \cite{Argyres:1996eh} and therefore identical when formulated in dimensions between 3 and 6. 
Another avenue for computing moduli spaces of instantons, where no such simplification is available, is through the \emph{Coulomb branch} of certain 3d gauge theories with 8 supercharges and gauge group  $\cG$ whose details we specify below. These two routes, via the Higgs branch and the Coulomb branch, are independent of each other, though calculating exact quantities on both sides can furnish a test of mirror symmetry and relate one to the other \cite{Intriligator:1996ex}. In this paper we will exclusively study theories whose Coulomb branch is the moduli space of instantons, without resorting to mirror symmetry.  

The stringy realization of moduli spaces of instantons through brane constructions has led to new insights. Indeed the ADHM construction exists only for classical gauge groups and, until recently, the instanton partition functions for exceptional gauge groups were only possible by means of superconformal indices \cite{Romelsberger:2005eg,Kinney:2005ej,Gadde:2011uv} of theories obtained by wrapping $M5$-branes on punctured Riemann surfaces \cite{Gaiotto:2009we} as in \cite{Gaiotto:2012uq} for $E_{6,7,8}$ instantons, by extrapolating the blow-up equations of \cite{Nakajima:2003pg,Nakajima:2005fg} as in \cite{Keller:2012da}, or by utilizing the generating function of holomorphic functions on the moduli space as in \cite{Benvenuti:2010pq, Keller:2011ek, Hanany:2012dm}. In this paper we explore the latter generating function, known as Hilbert series and shortened by HS, which counts gauge invariant chiral operators in a supersymmetric gauge theory \cite{Feng:2007ur}. We focus on supersymmetric gauge theories with $8$ supercharges whose moduli spaces include moduli spaces of instantons. 

Mathematically, the HS is a character of the global symmetry group of the ring of holomorphic functions on the moduli space of the supersymmetric gauge theory. It provides useful exact information about the moduli space: from the HS we can extract the group theoretic properties of the generators of the moduli space and of the relations between them. Salient features of the theories, such as the enhancement of global symmetries, are also neatly exposed by this treatment.
For moduli spaces of $k$ pure Yang-Mills instantons, the Hilbert series is also the five-dimensional (or K-theoretic) $k$ instanton partition function of \cite{Nekrasov:2002qd, Nakajima:2003pg, Nekrasov:2004vw, Nakajima:2005fg, Keller:2011ek}.

In \cite{Hanany:2012dm} the Hilbert series for instantons of charge $k=2$ were approached from the Higgs branch point of view, the calculations being a generalization of \cite{Benvenuti:2010pq} with an increased level of difficulty. Here we attack the problem from the Coulomb branch perspective in the wake of the new developments of \cite{Cremonesi:2013lqa}, where a simple formula for the Hilbert series of the Coulomb branch of $d=3$ $\cN=4$ \emph{good} or $\emph{ugly}$ \cite{Gaiotto:2008ak} superconformal field theories was introduced.%
\footnote{It was recently realized that the Coulomb branch Hilbert series of a $d=3$ $\cN=4$ theory is also captured by a limit of the superconformal index of the theory \cite{Razamat:2014pta}.}
The methods introduced in \cite{Cremonesi:2013lqa} have already given fruitful results \cite{Cremonesi:2014kwa,Cremonesi:2014vla}. 
Here we continue to exploit the techniques to analyze the moduli spaces of higher $k$ $G$-instantons, where $G$ is any simple Lie group. Our results include instantons for gauge groups whose Dynkin diagrams are non-simply laced, which have escaped a construction so far. 

The Coulomb branch of three-dimensional theories with $8$ supercharges receives quantum corrections and it is precisely this which begets the non-trivial structure of the space. As we will review in section \ref{sec:Hilbert}, the chiral operators which parametrize the Coulomb branch are gauge invariant combinations of supersymmetric 't Hooft monopole operators $V_m$ \cite{Borokhov:2002ib} labeled by a magnetic charge $m$, which break the gauge group $\cG$ to a subgroup $H_m$ by the adjoint Higgs mechanism, and of the classical complex scalar fields $\phi_m$ in the adjoint representation of the residual gauge group $H_m$. The HS of the Coulomb branch counts gauge invariant either bare (\emph{i.e.} built out of $V_m$ only) or dressed (\emph{i.e.} built out of $V_m$ and $\phi_m$) supersymmetric monopole operators according to their quantum numbers, namely the topological charges $J$ and the $R$-charge under the $U(1)_C$ Cartan subgroup of the $SU(2)_C$ $R$-symmetry which acts on the Coulomb branch.



Since we want to study moduli spaces of instantons we must make precise which theories, whose Coulomb branch we will investigate, are of interest to us. We extend the correspondence between the Coulomb branch of ADE quivers \cite{Kronheimer:1989zs,KronheimerNakajima} and the moduli space of ADE instantons, first pointed out for one instanton in \cite{Intriligator:1996ex} and then generalized to higher instanton number in \cite{deBoer:1996mp,Porrati:1996xi}. We claim that the moduli spaces of instantons for any simple gauge group can be obtained as the Coulomb branch of quivers constructed using the over-extension of the Dynkin diagrams for the associated finite Lie algebras. Whilst this has already been expounded using Hilbert series in \cite{Cremonesi:2013lqa,Cremonesi:2014vla} for ADE quivers, here we complete the treatment by generalizing the previous formula to non-simply laced quivers. The crucial formula that prescribes how to deal with multiple laces is \eqref{dimension_hyper_guess}.

The plan for the rest of this paper is as follows. Section \ref{sec:quiversAndBranes} is a brief summary of a particular type of brane construction that realizes instanton moduli spaces in string theory both from the Higgs branch and the Coulomb branch point of view.  From the brane picture we are able to motivate the quiver theories that we use to compute the Hilbert series of moduli spaces of instantons. In section \ref{sec:Hilbert} we 
review the monopole formula for the Hilbert series of Coulomb branches and we show how to modify the expression to account for generalized quivers built from non-simply laced Dynkin diagrams. In section \ref{sec:k-G2} we provide a step-by-step calculation for the moduli space of $k$ $G_2$ instantons and give the explicit result for the Hilbert series associated to the moduli space of 3 $G_2$ instantons. In sections \ref{sec:k-SO(2N+1)}, \ref{sec:k-USp(2N)}, \ref{sec:k-F4} we display formulae for the Hilbert series of $SO(2N+1)$, $USp(2N)$ and $F_4$ instantons. In section \ref{sec:modSpaceAlgebraicVariety} we sketch some of the group theoretic features of the moduli space of instantons as an algebraic variety, providing the transformation laws of the generators and the first relations. In section \ref{sec:concl} we present our conclusions.

\section{Brane realization of instantons} \label{sec:quiversAndBranes}
\emph{In this section we summarize various brane constructions for moduli spaces of instantons of classical gauge groups \cite{Douglas:1995bn,Witten:1995gx,deBoer:1996ck,Kapustin:1998fa,Hanany:1999sj,Hanany:2001iy}. String dualities which realize mirror symmetry relate the Higgs branch and the Coulomb branch brane picture. 
However we stress 
that the Coulomb branch construction that will be used later on does not require mirror symmetry. 
The mathematically oriented reader can skip this section altogether.} 
\newline

An instanton is a solitonic object of codimension 4. $Dp$-branes inside $D(p+4)$-branes, with or without $O(p+4)$-planes, provide a realization of instantons for classical gauge groups. To realize the kind of three-dimensional theory that we are interested in, we consider $D2$-branes in the background of $D6$-branes. The $D6$-branes provide the gauge group whilst $k$ $D2$-branes, when lying on top of the $D6$-branes, give rise to instanton configurations of charge $k$ on $\mathbb{C}^2$. The classical gauge group on the worldvolume of the $D6$-branes depends on which type of orientifold $O6$-plane is added to the construction. 
\bgroup
\def\arraystretch{1.5}%
\begin{table} [htdp]
\begin{center}
\begin{tabular}{|c|c|c|c|}
\hline
$G$ 
& \begin{tabular}[c]{@{}c@{}}Brane configurations from which \\ Higgs branch can be realized \end{tabular} & \hspace{2cm} ADHM quiver \hspace{2cm}  \\
\hline
$A_{N-1}$  
& \hspace{1.5cm} \begin{tikzpicture} [baseline=0]
\draw (0,0)--(0,1.5);
\draw (1.3,0)--(1.3,1.5); \draw (1.6,0)--(1.6,1.5); \draw (1.9,0)--(1.9,1.5);
\draw [decorate, decoration={brace, mirror}](0,-0.1) -- (1.9,-0.1) node[black,midway,yshift=-0.3cm] {\footnotesize $N~D6$};
\draw [loosely dotted] (0,0.75)--(1.3,0.75);
\fill (2.2,0.2) circle [radius=0.6mm];
\draw [loosely dotted] (2.2,0.3)--(2.2,0.9);
\fill (2.2,1) circle [radius=0.6mm];
\fill (2.2,1.2) circle [radius=0.6mm];
\node at (2.5,1.5) {\footnotesize $k$ $D2$};
\end{tikzpicture} 
&  \hspace{-2cm} 
\begin{tikzpicture}[font=\large, transform canvas={scale=0.6},baseline=-0.5cm]
\begin{scope}[auto,%
  every node/.style={draw, minimum size=1cm}, node distance=3cm];
\node [circle, label={[xshift=0.4cm, yshift=0cm] $U(k)$}] (Uk) at (0,0) {};
\node[rectangle, right=of Uk, label={[yshift=0cm] $SU(N)$}] (SUN) {};  
\end{scope}
\draw   (Uk)--(SUN) ;
\draw(-0.1,0.55) arc (30:325:0.95cm) node at (-2.5,0) {${\rm Adj}$};
\end{tikzpicture} \qquad \qquad \\
\hline
$B_{N}$  
& 
\begin{tikzpicture} [baseline=0]
\draw (0,0)--(0,1.5);
\draw (0.3,0)--(0.3,1.5); \draw (0.6,0)--(0.6,1.5); \draw (1.2,0)--(1.2,1.5); 
\draw [decorate, decoration={brace, mirror}](0,-0.2) -- (1.2,-0.2) node[black,midway,yshift=-0.3cm] {\footnotesize $N$ $D6$};
\draw [loosely dotted] (0.6,0.75)--(1.2,0.75);
\draw [dashed] (-0.3,0)--(-0.3,1.5);
\node at (-0.3,1.8) {\footnotesize $\tO 6^{-}$};
\begin{scope}[xscale=-1, xshift=0.6cm]
\draw (0,0)--(0,1.5);
\draw (0.3,0)--(0.3,1.5); \draw (0.6,0)--(0.6,1.5); \draw (1.2,0)--(1.2,1.5); 
\draw [decorate, decoration={brace}](0,-0.2) -- (1.2,-0.2) node[black,midway,yshift=-0.3cm] {\footnotesize $N$ $D6$ images};
\draw [loosely dotted] (0.6,0.75)--(1.2,0.75);
\end{scope}
\fill (1.5,0.2) circle [radius=0.6mm];
\draw [loosely dotted] (1.5,0.3)--(1.5,0.9);
\fill (1.5,1) circle [radius=0.6mm];
\fill (1.5,1.2) circle [radius=0.6mm];
\node at (1.7,1.5) {\footnotesize $k$ $D2$};
\fill (-2.1,0.2) circle [radius=0.6mm];
\draw [loosely dotted] (1.5,0.3)--(1.5,0.9);
\fill (-2.1,1) circle [radius=0.6mm];
\fill (-2.1,1.2) circle [radius=0.6mm];
\node at (-2.9,1.5) {\footnotesize $k$ $D2$ images};
\end{tikzpicture}  &  \hspace{-2cm}
\begin{tikzpicture}[font=\large, transform canvas={scale=0.6}]
\begin{scope}[auto,%
  every node/.style={draw, minimum size=1cm}, node distance=3cm];
\node [circle, label={[xshift=0.7cm, yshift=0cm] $USp'(2k)$}] (Uk) at (0,0) {};
\node[rectangle, right=of Uk, label={[yshift=0cm] $SO(2N+1)$}] (SUN) {};  
\end{scope}
\draw   (Uk)--(SUN) ;
\draw(-0.1,0.55) arc (30:325:0.95cm) node at (-2.3,0) {${\rm A}$};
\end{tikzpicture} \qquad \qquad \\
\hline
$C_{N}$  
& \begin{tikzpicture} [baseline=0]
\draw (0,0)--(0,1.5);
\draw (0.3,0)--(0.3,1.5); \draw (0.6,0)--(0.6,1.5); \draw (1.2,0)--(1.2,1.5); 
\draw [decorate, decoration={brace, mirror}](0,-0.2) -- (1.2,-0.2) node[black,midway,yshift=-0.3cm] {\footnotesize $N$ $D6$};
\draw [loosely dotted] (0.6,0.75)--(1.2,0.75);
\draw [dashed] (-0.3,0)--(-0.3,1.5);
\node at (-0.3,1.8) {\footnotesize  $O6^{+}$};
\begin{scope}[xscale=-1, xshift=0.6cm]
\draw (0,0)--(0,1.5);
\draw (0.3,0)--(0.3,1.5); \draw (0.6,0)--(0.6,1.5); \draw (1.2,0)--(1.2,1.5); 
\draw [decorate, decoration={brace}](0,-0.2) -- (1.2,-0.2) node[black,midway,yshift=-0.3cm] {\footnotesize $N$ $D6$ images};
\draw [loosely dotted] (0.6,0.75)--(1.2,0.75);
\end{scope}
\fill (1.5,0.2) circle [radius=0.6mm];
\draw [loosely dotted] (1.5,0.3)--(1.5,0.9);
\fill (1.5,1) circle [radius=0.6mm];
\fill (1.5,1.2) circle [radius=0.6mm];
\node at (1.7,1.5) {\footnotesize $k$ $D2$};
\fill (-2.1,0.2) circle [radius=0.6mm];
\draw [loosely dotted] (1.5,0.3)--(1.5,0.9);
\fill (-2.1,1) circle [radius=0.6mm];
\fill (-2.1,1.2) circle [radius=0.6mm];
\node at (-2.9,1.5) {\footnotesize $k$ $D2$ images};
\end{tikzpicture} 
&   \hspace{-2cm}
\begin{tikzpicture}[font=\large, transform canvas={scale=0.6}]
\begin{scope}[auto,%
  every node/.style={draw, minimum size=1cm}, node distance=3cm];
\node [circle, label={[xshift=0.7cm, yshift=0cm] $O(2k)$}] (Ok) at (0,0) {};
\node[rectangle, right=of Ok, label={[yshift=0cm] $USp(2N)$}] (USp2N) {};  
\end{scope}
\draw   (Ok)--(USp2N) ;
\draw(-0.1,0.55) arc (30:325:0.95cm) node at (-2.3,0) {${\rm S}$};
\end{tikzpicture} \qquad \qquad \\
\hline
$C_{N}$  
& \begin{tikzpicture} [baseline=0]
\draw (0,0)--(0,1.5);
\draw (0.3,0)--(0.3,1.5); \draw (0.6,0)--(0.6,1.5); \draw (1.2,0)--(1.2,1.5); 
\draw [decorate, decoration={brace, mirror}](0,-0.2) -- (1.2,-0.2) node[black,midway,yshift=-0.3cm] {\footnotesize $N$ $D6$};
\draw [loosely dotted] (0.6,0.75)--(1.2,0.75);
\draw [dashed] (-0.3,0)--(-0.3,1.5);
\node at (-0.3,1.8) {\footnotesize  $\tilde{O}6^{+}$};
\fill (-0.3,0.9) circle [radius=0.6mm];
\begin{scope}[xscale=-1, xshift=0.6cm]
\draw (0,0)--(0,1.5);
\draw (0.3,0)--(0.3,1.5); \draw (0.6,0)--(0.6,1.5); \draw (1.2,0)--(1.2,1.5); 
\draw [decorate, decoration={brace}](0,-0.2) -- (1.2,-0.2) node[black,midway,yshift=-0.3cm] {\footnotesize $N$ $D6$ images};
\draw [loosely dotted] (0.6,0.75)--(1.2,0.75);
\end{scope}
\fill (1.5,0.2) circle [radius=0.6mm];
\draw [loosely dotted] (1.5,0.3)--(1.5,0.9);
\fill (1.5,1) circle [radius=0.6mm];
\fill (1.5,1.2) circle [radius=0.6mm];
\node at (1.7,1.5) {\footnotesize $k$ $D2$};
\fill (-2.1,0.2) circle [radius=0.6mm];
\draw [loosely dotted] (1.5,0.3)--(1.5,0.9);
\fill (-2.1,1) circle [radius=0.6mm];
\fill (-2.1,1.2) circle [radius=0.6mm];
\node at (-2.9,1.5) {\footnotesize $k$ $D2$ images};
\end{tikzpicture} 
&   \hspace{-2cm}
\begin{tikzpicture}[font=\large, transform canvas={scale=0.6}]
\begin{scope}[auto,%
  every node/.style={draw, minimum size=1cm}, node distance=3cm];
\node [circle, label={[xshift=0.7cm, yshift=0cm] $O(2k+1)$}] (Ok) at (0,0) {};
\node[rectangle, right=of Ok, label={[yshift=0cm] $USp'(2N)$}] (USp2N) {};  
\end{scope}
\draw   (Ok)--(USp2N) ;
\draw(-0.1,0.55) arc (30:325:0.95cm) node at (-2.3,0) {${\rm S}$};
\end{tikzpicture} \qquad \qquad \\
\hline
$D_{N}$  
& \begin{tikzpicture} [baseline=0]
\draw (0,0)--(0,1.5);
\draw (0.3,0)--(0.3,1.5); \draw (0.6,0)--(0.6,1.5); \draw (1.2,0)--(1.2,1.5); 
\draw [decorate, decoration={brace, mirror}](0,-0.2) -- (1.2,-0.2) node[black,midway,yshift=-0.3cm] {\footnotesize $N$ $D6$};
\draw [loosely dotted] (0.6,0.75)--(1.2,0.75);
\draw [dashed] (-0.3,0)--(-0.3,1.5);
\node at (-0.3,1.8) {\footnotesize $O6^{-}$};
\begin{scope}[xscale=-1, xshift=0.6cm]
\draw (0,0)--(0,1.5);
\draw (0.3,0)--(0.3,1.5); \draw (0.6,0)--(0.6,1.5); \draw (1.2,0)--(1.2,1.5); 
\draw [decorate, decoration={brace}](0,-0.2) -- (1.2,-0.2) node[black,midway,yshift=-0.3cm] {\footnotesize $N$ $D6$ images};
\draw [loosely dotted] (0.6,0.75)--(1.2,0.75);
\end{scope}
\fill (1.5,0.2) circle [radius=0.6mm];
\draw [loosely dotted] (1.5,0.3)--(1.5,0.9);
\fill (1.5,1) circle [radius=0.6mm];
\fill (1.5,1.2) circle [radius=0.6mm];
\node at (1.7,1.5) {\footnotesize $k$ $D2$};
\fill (-2.1,0.2) circle [radius=0.6mm];
\draw [loosely dotted] (1.5,0.3)--(1.5,0.9);
\fill (-2.1,1) circle [radius=0.6mm];
\fill (-2.1,1.2) circle [radius=0.6mm];
\node at (-2.9,1.5) {\footnotesize $k$ $D2$ images};
\end{tikzpicture} 
&   \hspace{-2cm}
\begin{tikzpicture}[font=\large, transform canvas={scale=0.6}]
\begin{scope}[auto,%
  every node/.style={draw, minimum size=1cm}, node distance=3cm];
\node [circle, label={[xshift=0.7cm, yshift=0cm] $USp(2k)$}] (USp2k) at (0,0) {};
\node[rectangle, right=of USp2k, label={[yshift=0cm] $SO(2N)$}] (SO2N) {};  
\end{scope}
\draw   (USp2k)--(SO2N) ;
\draw(-0.1,0.55) arc (30:325:0.95cm) node at (-2.3,0) {${\rm A}$};
\end{tikzpicture} \qquad \qquad \\
\hline
\end{tabular}
\end{center}
\caption{Brane constructions and quiver diagrams whose Higgs branch correspond to $k$ $G$-instantons on $\BC^2$.
To describe the moduli space of instantons, all D2 branes are dissolved on coincident D6 branes and orientifold planes. In the pictures the $D6$ branes are separated from each other and the orientifold for clarity. Note that there exist constructions of the moduli space of E-instantons in terms of M5-branes on a sphere with punctures. However it is unknown how to realize such moduli spaces as perturbative open string backgrounds.}
\label{tab:instantonBraneConstHiggsBranch}
\end{table}%
In particular $N$ parallel $D6$-branes provide a $U(N)$ low energy effective theory, as sketched in \tref{tab:instantonBraneConstHiggsBranch}. With the addition of $k$ $D2$-branes, the system living on the latter becomes that of a quiver theory with gauge group $U(k)$ and $SU(N)$ flavor symmetry, since the $U(1)$ factor inside $U(N)$ is gauged.

In order to realize $SO(2N+1)$ instantons we construct a background with $N$ parallel $D6$-branes on top of an orientifold plane $\widetilde{O}6^{-}$. The orientifold allows for strings to end on it, thus reproducing the $B_N$ root system. The quiver for such a construction is given by a gauge group $USp(2k)$ with matter in the antisymmetric representation and $2N+1$ fundamental half-hypermultiplets with flavor symmetry $SO(2N+1)$.%
\footnote{We have glossed over a subtlety: the $\widetilde{O}6^{-}$ plane requires the presence of a Romans mass. This $D8$-brane charge translates into a Chern-Simons coupling in the parity anomalous gauge theory on $D2$-branes, which reduces supersymmetry and lifts the Coulomb branch. The moduli space of $B_N$ instantons is the subvariety of the total moduli space of vacua of the supersymmetric Chern-Simons theory with vanishing expectation values for monopole operators.}

For $USp(2N)$-instantons, the brane construction involves $N$ $D6$-branes on top of an $O6^{+}$ or $\widetilde{O}6^{+}$ plane. $k$ half $D2$-branes in such a background give rise to a quiver gauge theory with $O(k)$ gauge group, matter in the symmetric representation and $2N$ fundamental half-hypermultiplets with flavor symmetry $USp(2N)$.

Lastly, in presence of  $k$ $D2$-branes, $N$ $D6$-branes and an orientifold $O6^{-}$, the $D_N$ root system is realized, allowing for a quiver with $USp(2k)$ gauge symmetry, matter in the antisymmetric representation and $2N$ fundamental half-hypermultiplets with flavor symmetry $SO(2N)$.

The Higgs branch of these theories is achieved when the $D2$-branes are inside the $D6$-branes; the Coulomb branch is realized when the $D2$-branes are away from the $D6$-branes. Thus it is the Higgs branch of these quiver gauge theories that reproduces the moduli space of $G$-instantons, where $G$ is the flavor symmetry group of the quiver. We show the brane constructions and the corresponding quivers in \tref{tab:instantonBraneConstHiggsBranch}. 

For exceptional groups we do not have a perturbative open string description on the Higgs branch. However progress can be made appealing to mirror symmetry and generalizing the lessons learned for classical groups. We can implement this symmetry on the previous constructions by performing $T$-duality to Type IIB, and then $S$-duality to realize mirror symmetry. Under T-duality along a direction transverse to the $D2$-branes and parallel to the $D6$-branes, the $D2$-brane becomes a $D3$-brane on $S^1$ and the $D6$-brane becomes a $D5$-brane.%
\footnote{More precisely, we view $\bC^2=\bR^4$ as an ``$A_0$'' hyperK\"ahler space, namely a circle fibration over $\bR^3$ with a fixed point, and perform $T$-duality along the fiber. The fixed point of the circle action is dualized to an $NS5$-brane. We will return to this point in the following.\label{footnoteTdualityNS}}
After $S$-duality, the $D3$-brane is unchanged whilst the $D5$-brane turns into a $NS5$-brane. In the absence of orientifolds, i.e. for the case of $G=A_{N-1}$ in \tref{tab:instantonBraneConstCoulombBranch1}, the application of these dualities results in a necklace quiver gauge theory with $N$ $U(k)$ gauge nodes.

Moreover, and crucially, since mirror symmetry exchanges Higgs branches with Coulomb branches, it is now the Coulomb branch of this new dual theory which corresponds to the moduli space of instantons.

The action of mirror symmetry on the four orientifold planes we considered is illustrated in ~\tref{tab:TandSonOrientifolds}. Note in particular that $T$-duality results in a restriction to an interval defined by two separated $O5$ planes and that $S$-duality turns an $O5$ into an $ON$ orientifold.

\begin{table}[htdp]
\begin{center}
\begin{tabular} {|c|c|c|}
\hline
Orientifold & $T$-duality & $S$-duality \\
\hline
$\widetilde{O6}^-$ & $O5^-$ \& $\widetilde{O}5^-$ & $ON^-$ \& $\widetilde{ON}^-$ \\
$O6^+$ & $O5^+$ \& $O5^+$ & $ON^+$ \& $ON^+$ \\
$O6^-$ & $O5^-$ \& $O5^-$ & $ON^-$ \& $ON^-$ \\
$\widetilde{O}6^+$ & $\widetilde{O}5^+$ \& $O5^+$ & $\widetilde{O}N^+$ \& $ON^+$ \\
\hline
\end{tabular}
\end{center}
\caption{The effect of $T$- and $S$-dualities on orientifold planes.}
\label{tab:TandSonOrientifolds}
\end{table}%

The effect of mirror symmetry, through action on branes and orientifolds, on the brane constructions in ~\tref{tab:instantonBraneConstHiggsBranch} is summarized in ~\tref{tab:instantonBraneConstCoulombBranch1}. For example consider the brane realization on the Higgs branch of one $C_N$ instanton (i.e with $k=1$ $D2$-branes). The $O6^+$ background is turned into an interval bounded by $ON^+$ on the left and an $ON^+$ on the right. The $N$ parallel $NS5$-branes lie within this interval. 

As befits a magnetically charged object, the $D3$-brane is to be viewed as a root of the Langlands dual algebra, here $B_N$. When stretching onto the $ON^+$, the $D3$-brane reproduces a short root: it ends on the $ON^+$. Finally, one balances the number of $D3$-branes stretching between neighboring $NS5$, in this case one. The result is sketched in \fref{fig:BraneCn}.

After engineering the dual brane construction, we can associate to it a quiver.
The rank of each node in the quiver is read off from the number of $D$-branes: since we have one $D3$-brane between each neighboring $NS5$, the gauge groups are all $U(1)$.

To account for the different length of the last root on the left and on the right, we use the double lace notation of Dynkin diagrams. In the next section we will specify how to deal with multiple laces. The quiver we end up with is the \emph{Dynkin diagram of the untwisted affine algebra $C^{(1)}_N$}, with the dual Coxeter labels (or Kac labels/comarks) $a_i^\vee$, $i=0,\dots, r=\mathrm{rk}(G)$,  providing the rank of the gauge groups. For instanton number $k$ the ranks of the unitary gauge groups are given by $k a_i^\vee$. 

\begin{figure}[H]
\centering
\begin{tikzpicture} [baseline=0]
\draw [ultra thick] (0,0)--(0,2.5) node[black,midway, xshift =-0.3cm, yshift=-1.5cm] {\footnotesize $ON^+$} node[midway, color=blue, xshift=0.2cm, yshift=0.4cm] {$\bullet$} node[black,midway, xshift=0.25cm, yshift=0.7cm] {\scriptsize D5};;
\draw (0.5,0)--(0.5,2.5); \draw (1,0)--(1,2.5); \draw (1.5,0)--(1.5,2.5); \draw (3,0)--(3,2.5); \draw (3.5,0)--(3.5,2.5); \draw (4,0)--(4,2.5); 
\draw [ultra thick] (4.5,0)--(4.5,2.5) node[black,midway, xshift =0.3cm, yshift=-1.5cm] {\footnotesize $ON^+$};
%
\draw [thick, color=red](0.02,1)--(0.5,1) node[black,midway, yshift=-0.2cm] {} ;
%
\draw (0.5,0.9)--(1,0.9) node[black,midway, yshift=0.2cm] {};
\draw (1,1.3)--(1.5,1.3) node[black,midway, yshift=0.2cm] {} node[black,midway, yshift=-0.3cm] {\tiny D3} ;
\draw [loosely dotted] (1.5,1)--(3,1);
\draw (3,1.3)--(3.5,1.3) node[black,midway, yshift=0.2cm] {} node[black,midway, xshift =0.3cm, yshift=1.4cm] {\footnotesize NS5};
\draw (3.5,1.7)--(4,1.7) node[black,midway, yshift=0.2cm] {};
%
\draw [thick, color=red](4,0.75)--(4.48,0.75) node[black,midway, yshift=0.2cm] {};
%
\draw [decorate, decoration={brace, mirror}](0.5,-0.1)--(4,-0.1) node[black,midway,yshift=-0.5cm] {\footnotesize $N-1~\text{intervals}$};
\end{tikzpicture}  
\caption{Brane construction for the $C_N$ affine Dynkin diagram with the attached $U(1)$ node. Each type of brane is indicated in the diagram. Here there is one $D3$-brane per interval. The red and black segments indicate D3-branes in correspondence with the simple roots of the $B$-type algebra, which is dual to the $C$-type algebra associated with $ON^+$. The blue dot in the leftmost interval indicates the D5-brane corresponding to the over-extended $U(1)$ node.}
\label{fig:BraneCn}
\end{figure}
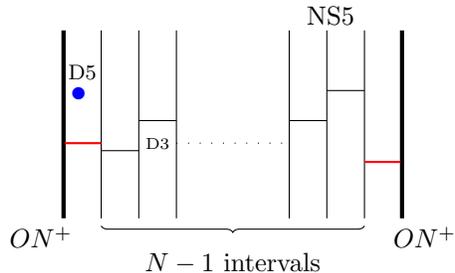

In a completely analogous fashion to this example, the quivers that we analyze for the moduli space of $G$-instantons are precisely the Dynkin diagrams for the untwisted affine algebras of $G$ type, with the crucial addition of an extra node, the nature of which we explain below.%
\footnote{We have chosen to use the untwisted affine Dynkin diagrams associated to \emph{electric} objects, rather than the Langlands dual Dynkin diagrams associated to \emph{magnetic objects}, which are obtained by reversing the arrows. The prescription that we will provide for the HS of instanton moduli spaces from the Coulomb branch can be phrased equally well in terms of dual diagrams.}

\paragraph{Over-extended node} 

The quiver gauge theories constructed from the affine Dynkin diagrams are not sufficient to obtain the moduli spaces of instantons. In particular, for $k>1$ instanton number, the parametrization of the instanton solution on $\mathbb{C}^2$ mixes nontrivially with the parametrization of the instanton in the gauge group $G$. 

For $k=1$, i.e. a single $D3$ brane stretching on a circle, the fugacity associated with $\mathbb{C}^2$ factorizes:
\bea \label{factorisableModSpace}
g_{1,G}(t,x,\vec{u})=\frac{1}{(1-tx)(1-tx^{-1})} {\tilde g}_{1,G}(t,\vec{u})~.
\eea
Here $\vec{u}$ are the fugacities associated to $G$,%
\footnote{In this paper we use simple roots fugacities $\vec{u}$ instead of highest weight fugacities $\vec{y}$ for convenience.} $x$ is the fugacity associated to $SU(2)$ rotations of $\mathbb{C}^2$, and $t$ the fugacity for the highest weight of the $SU(2)$ $R$-symmetry. After factoring out the center of mass degree of freedom, we are left with the Hilbert series $ {\tilde g}_{1,G}$ of the reduced moduli space of $1$ $G$-instanton, which does not depend on $x$.

For $k>1$ one can similarly extract the center of mass mode, 
\bea \label{nonfactorisableModSpace}
g_{k,G}(t,x,\vec{u})=\frac{1}{(1-tx)(1-tx^{-1})} {\tilde g}_{k,G}(t,x,\vec{u})~,
\eea
but the Hilbert series ${\tilde g}_{k,G}$ of the reduced moduli space of $k$ $G$-instantons depends on the $SU(2)$ fugacity $x$ for $k>1$. In fact, as we will explain in section \ref{sec:mon_global_symm}, for $k>1$ there are two different global $SU(2)$ symmetries, one acting on the center of mass and the other on the reduced moduli space of instantons.

In order to see the center of mass of the instantons and the $SU(2)_x$ rotation symmetry of $\bC^2$ in the Type IIB brane construction, we need to follow the chain of dualities more carefully (see footnote \ref{footnoteTdualityNS}). 
The $T$-duality from Type IIA to Type IIB is done along a circle direction with a fixed point: this results in an extra $NS5$ brane in Type IIB, in addition to the $D5$-branes and $O5$-planes discussed above. The $NS5$-brane ensures that the matter fields in the $2$-index tensor representation of the ADHM quiver gauge groups transform as denoted in Table \ref{tab:instantonBraneConstHiggsBranch} rather than the adjoint representation. $S$-duality maps this $NS5$-brane into a $D5$-brane, which fixes the origin of $\mathbb{C}^2$. The $D5$-brane $U(1)$ symmetry acts as a flavor group for the worldvolume theory on the $D3$-branes:  it attaches a square node to the extended node of the affine Dynkin diagram, as in \cite{deBoer:1996mp, Porrati:1996xi, deBoer:1996ck}. 

Even though this $U(1)$ node appears naturally as a flavor node in the brane construction, it is useful to treat it on the same footing as the other gauge nodes, and then ungauge an overall diagonal $U(1)$ gauge symmetry under which no matter fields are charged. The relevant quivers for the moduli space of instantons on $\mathbb{C}^2$ are then the so-called \emph{over-extended} Dynkin diagrams \cite{Julia:1982gx}, with a rank $1$ over-extended node connected to the extended (or affine) node. 
The gauge fixing of the decoupled $U(1)$ gauge symmetry can be done at any node of the quiver: fixing the $U(1)$ of the over-extended node reduces it to a flavor node as is natural in the brane construction; fixing a $U(1)$ inside a $U(N)$ gauge factor leaves an $SU(N)/\bZ_N$ gauge group. In section \ref{sec:Hilbert} we explain how to implement this gauge fixing and how to identify the global symmetries acting on the instanton moduli space in the Coulomb branch Hilbert series.
\begin{table}[H] 
\begin{center}
{\footnotesize
\begin{tabular}{|c|c|c|c|}
\hline
$G$ & $\mathcal{L}$ & Coulomb branch quivers & Brane set-up\\
\hline
$A_{N}$ & Y & $\begin{array}{l} 
\raisebox{-12pt}{\rotatebox{30}{$-\!\!-\!\!-$}}\node{}{k}\raisebox{0pt}{\rotatebox{-30}{$-\!\!-\!\!-$}} \\[-7pt]
\node{}{k}-\node{}{k}\cdots-\Node{}{k} -{\blue \node{}{1}} 
\end{array}$ &  \begin{tabular}[c]{@{}c@{}} \\ \begin{tikzpicture} [scale=0.9, transform shape]
\draw (0,0.4)--(0,2);
\draw (1,0)--(1,1.6);
\draw (2,0)--(2,1.6);
\draw (3,0)--(3,1.6);
\draw (4,0.4)--(4,2) node[black,midway, yshift=1.2cm] {\footnotesize NS5} ;
\draw (0.7,0.9)--(0.7,2.5); 
\draw (1.7,0.9)--(1.7,2.5); 
%
\draw (0,0.9)--(1,0.5) node[black,midway, yshift=0.2cm] {\scriptsize $k$};
\draw (1,0.6)--(2,0.6) node[black,midway, yshift=0.2cm] {\scriptsize $k$};
\draw (2,0.5)--(3,0.5) node[black,midway, yshift=0.2cm] {\scriptsize $k$};
\draw (3,0.6)--(4,0.9) node[black,midway, yshift=0.2cm] {\scriptsize $k$} node[black,midway, xshift=-0.1cm, yshift=-0.3cm] {\scriptsize D3} ;
\draw (0,1.2)--(0.7,1.5) node[black,midway, yshift=0.2cm] {\scriptsize $k$};
\draw (0.7,1.7)--(1.7,1.7) node[black,midway, xshift=0.2cm, yshift=-0.2cm] {\scriptsize $k$} node[midway, color=blue, yshift=0.5cm] {$\bullet$} node[black,midway, yshift=0.8cm] {\scriptsize D5};
\draw [loosely dotted,rounded corners=0.95cm] (1.7,1.8)--(3.2,1.8)--(4,1.4);
%
\draw [decorate, decoration={brace, mirror}](0,-0.1)--(4,-0.1) node[black,midway,yshift=-0.5cm] {\footnotesize $N~\text{intervals}$};
\end{tikzpicture} \end{tabular} \\
$B_{N}$ & N  & ${\blue \node{}{1}}- \Node{}{k}-\node{\ver{}{k}}{2k}-\underbrace{\node{}{2k}-\cdots-\node{}{2k}}_{N-3~\text{nodes}}\Rightarrow\node{}{k}  \ \tikz[na]\node(B1){}; $ &  \begin{tabular}[c]{@{}c@{}} \\  \begin{tikzpicture} [baseline=0, scale=0.9, transform shape]
\draw [ultra thick] (0,0)--(0,2.5) node[black,midway, xshift =-0.3cm, yshift=-1.5cm] {\footnotesize $ON^-$} node[midway, color=blue, xshift=0.2cm, yshift=0.4cm] {$\bullet$} node[black,midway, xshift=0.25cm, yshift=0.7cm] {\scriptsize D5};;
\draw (0.5,0)--(0.5,2.5); \draw (1,0)--(1,2.5); \draw (1.5,0)--(1.5,2.5); \draw (3,0)--(3,2.5); \draw (3.5,0)--(3.5,2.5); \draw (4,0)--(4,2.5); 
\draw [ultra thick] (4.5,0)--(4.5,2.5) node[black,midway, xshift =0.3cm, yshift=-1.5cm] {\footnotesize $\widetilde{O}N^-$};
%
\draw [thick, color=red, rounded corners=0.75cm](0.5,1)--(-0.4,0.75)--(1,0.75) node[black,midway, xshift=-0.0cm, yshift=-0.2cm] {\scriptsize $k$} ;
%
\draw (0.5,0.9)--(1,0.9) node[black,midway, yshift=0.4cm] {\scriptsize $k$};
\draw (1,1.3)--(1.5,1.3) node[black,midway, yshift=0.4cm] {\scriptsize $2k$} ;
\draw (1,1.5)--(1.5,1.5) node[black,midway, yshift=-0.4cm] {\tiny D3} ;
\draw [loosely dotted] (1.5,1)--(3,1);
\draw (3,1.3)--(3.5,1.3) node[black,midway, yshift=0.4cm] {\scriptsize $2k$};
\draw (3,1.5)--(3.5,1.5) node[black,midway, xshift =0.3cm, yshift=1.4cm] {\footnotesize NS5} ;
\draw (3.5,1.7)--(4,1.7) node[black,midway, yshift=0.5cm] {\scriptsize $2k$};
\draw (3.5,1.9)--(4,1.9);
%
\draw [thick, color=red, rounded corners=0.75cm](4,0.75)--(4.9,0.75) --(4,1) node[black,midway, xshift=-0.2cm, yshift=0.3cm] {\scriptsize $k$};
%
\draw [decorate, decoration={brace, mirror}](1,-0.1)--(4,-0.1) node[black,midway,yshift=-0.5cm] {\footnotesize $N-2~\text{intervals}$};
\end{tikzpicture} \end{tabular}  \\
$C_{N}$ & N  & ${\blue \node{}{1}} -\Node{}{k}\Rightarrow\underbrace{ \node{}{k}-\cdots-\node{}{k}}_{N-1~\text{nodes}}\Leftarrow\node{}{k} \ \tikz[na]\node(C1){};$ &  \begin{tabular}[c]{@{}c@{}} \\ 
\begin{tikzpicture} [baseline=0, scale=0.9, transform shape]
\draw [ultra thick] (0,0)--(0,2.5) node[black,midway, xshift =-0.3cm, yshift=-1.5cm] {\footnotesize $ON^+$} node[midway, color=blue, xshift=0.2cm, yshift=0.4cm] {$\bullet$} node[black,midway, xshift=0.25cm, yshift=0.7cm] {\scriptsize D5};;
\draw (0.5,0)--(0.5,2.5); \draw (1,0)--(1,2.5); \draw (1.5,0)--(1.5,2.5); \draw (3,0)--(3,2.5); \draw (3.5,0)--(3.5,2.5); \draw (4,0)--(4,2.5); 
\draw [ultra thick] (4.5,0)--(4.5,2.5) node[black,midway, xshift =0.3cm, yshift=-1.5cm] {\footnotesize $ON^+$};
%
\draw [thick, color=red](0.02,1)--(0.5,1) node[black,midway, yshift=-0.2cm] {\scriptsize $k$} ;
%
\draw (0.5,0.9)--(1,0.9) node[black,midway, yshift=0.2cm] {\scriptsize $k$};
\draw (1,1.3)--(1.5,1.3) node[black,midway, yshift=0.2cm] {\scriptsize $k$} node[black,midway, yshift=-0.3cm] {\tiny D3} ;
\draw [loosely dotted] (1.5,1)--(3,1);
\draw (3,1.3)--(3.5,1.3) node[black,midway, yshift=0.2cm] {\scriptsize $k$} node[black,midway, xshift =0.3cm, yshift=1.4cm] {\footnotesize NS5};
\draw (3.5,1.7)--(4,1.7) node[black,midway, yshift=0.2cm] {\scriptsize $k$};
%
\draw [thick, color=red](4,0.75)--(4.48,0.75) node[black,midway, yshift=0.2cm] {\scriptsize $k$};
%
\draw [decorate, decoration={brace, mirror}](0.5,-0.1)--(4,-0.1) node[black,midway,yshift=-0.5cm] {\footnotesize $N-1~\text{intervals}$};
\end{tikzpicture}  \end{tabular} \\
$D_{N}$ & Y & $\node{}{k}-\node{\ver{}{k}}{2k}-\underbrace{\node{}{2k}-\cdots-\node{}{2k}}_{N-5~\text{nodes}}-\node{\ver{}{k}}{2k}-\Node{}{k}-{\blue \node{}{1}}$ & \begin{tabular}[c]{@{}c@{}} \\  \begin{tikzpicture} [baseline=0, scale=0.9, transform shape]
\draw [ultra thick] (0,0)--(0,2.5) node[black,midway, xshift =-0.3cm, yshift=-1.5cm] {\footnotesize $ON^-$} node[midway, color=blue, xshift=0.2cm, yshift=0.4cm] {$\bullet$} node[black,midway, xshift=0.25cm, yshift=0.7cm] {\scriptsize D5};;
\draw (0.5,0)--(0.5,2.5); \draw (1,0)--(1,2.5); \draw (1.5,0)--(1.5,2.5); \draw (3,0)--(3,2.5); \draw (3.5,0)--(3.5,2.5); \draw (4,0)--(4,2.5); 
\draw [ultra thick] (4.5,0)--(4.5,2.5) node[black,midway, xshift =0.3cm, yshift=-1.5cm] {\footnotesize $ON^-$};
%
\draw [thick, color=red, rounded corners=0.75cm](0.5,1)--(-0.4,0.75)--(1,0.75) node[black,midway, xshift=-0.0cm, yshift=-0.2cm] {\scriptsize $k$} ;
%
\draw (0.5,0.9)--(1,0.9) node[black,midway, yshift=0.4cm] {\scriptsize $k$};
\draw (1,1.3)--(1.5,1.3) node[black,midway, yshift=0.4cm] {\scriptsize $2k$} ;
\draw (1,1.5)--(1.5,1.5) node[black,midway, yshift=-0.4cm] {\tiny D3} ;
\draw [loosely dotted] (1.5,1)--(3,1);
\draw (3,1.3)--(3.5,1.3) node[black,midway, yshift=0.4cm] {\scriptsize $2k$};
\draw (3,1.5)--(3.5,1.5) node[black,midway, xshift =0.3cm, yshift=1.4cm] {\footnotesize NS5} ;
\draw (3.5,1.7)--(4,1.7) node[black,midway, yshift=0.5cm] {\scriptsize $k$};
%
\draw [thick, color=red, rounded corners=0.75cm](3.5,0.75)--(4.9,0.75) --(4,1) node[black,midway, xshift=-0.2cm, yshift=0.3cm] {\scriptsize $k$};
%
\draw [decorate, decoration={brace, mirror}](1,-0.1)--(3.5,-0.1) node[black,midway,yshift=-0.5cm] {\footnotesize $N-3~\text{intervals}$};
\end{tikzpicture}  \end{tabular} \\
\hline
\end{tabular}}
\end{center}
\caption{Quiver diagrams from which the Hilbert series of the moduli space of $k$ instanton in classical gauge groups can be computed using the monopole formula for the Coulomb branch. The corresponding brane configuration is depicted next to each quiver. Note that the configurations associated with the left boundary condition for $B_N$ and the left and right boundary conditions for $D_N$ involve an $ON^-$ plane and an NS5 brane, whose combination is usually called $ON^0$ \cite{Hanany:1999sj}; this type of configuration was pointed out in \cite{Sen:1998ii, Kapustin:1998fa}. The second column indicates whether a Lagrangian is available or not.}
\label{tab:instantonBraneConstCoulombBranch1}
\end{table}%
\begin{table}[H] 
\begin{center}
{\footnotesize
\begin{tabular}{|c|c|c|}
\hline
$G$ & $\mathcal{L}$ & Coulomb branch quivers \\ 
\hline
$E_{6}$ & Y  & ${\blue \node{}{1}} -\Node{}{k}-\node{}{2k}-\node{\overset{\ver{}{k}}{\ver{}{2k}}}{3k}-\node{}{2k}-\node{}{k}  \ \tikz[na];$  \\
$E_{7}$ & Y  & ${\blue \node{}{1}} -\Node{}{k}-\node{}{2k}-\node{}{3k}-\node{\ver{}{2k}}{4k}-\node{}{3k}-\node{}{2k}-\node{}{k} \ \tikz[na]; $  \\
$E_{8}$ & Y  & ${\blue \node{}{1}} -\Node{}{k}-\node{}{2k}-\node{}{3k}-\node{}{4k}-\node{}{5k}-\node{\ver{}{3k}}{6k}-\node{}{4k}-\node{}{2k}  \ \tikz[na]; $  \\
$F_4$   & N  & ${\blue \node{}{1}} -\Node{}{k}-\node{}{2k}-\node{}{3k}\Rightarrow\node{}{2k}-\node{}{k} \quad\tikz[na]\node(F41){}; $  \\
$G_2$   & N  & ${\blue \node{}{1}} -\Node{}{k}-\node{}{2k}\Rrightarrow\node{}{k} \quad\tikz[na]\node(G21){}; $ \\
\hline
\end{tabular}}
\end{center}
\caption{Quiver diagrams from which the Hilbert series of the moduli space of $k$ instantons in exceptional gauge groups can be computed using the monopole formula for the Coulomb branch. For these cases there is no known brane construction analogous to \tref{tab:instantonBraneConstCoulombBranch1}.}
\label{tab:instantonBraneConstCoulombBranch2}
\end{table}%


\section{The Hilbert series for the moduli space of $k$ $G$-instantons}\label{sec:Hilbert}

The purpose of this section is to review the essential tools for the computation of the Hilbert series for the quantum corrected Coulomb branch of $3d$ $\mathcal{N}=4$ quiver gauge theories where the gauge group is a product of $U(N)$ factors. As we have detailed in the previous section, for suitable generalized quivers, possibly including non-simple laces, this method provides the Hilbert series of the moduli space of instantons.

Three-dimensional $\mathcal{N}=4$ theories are described by vector multiplets in the adjoint representation and matter fields (hypermultiplets or half-hypermultiplets) transforming in some representation of the gauge group. At a generic point on the Coulomb branch the scalars in the vector multiplet acquire non-zero VEV, breaking the gauge group $\cG$ of rank $r$ to $U(1)^r$, its maximal torus; matter fields and W-bosons acquire mass and are integrated out, while the $r$ massless gauge fields, the photons, can be dualized to scalars.
So at low energies on the Coulomb branch, what is left is an effective theory of $r$ abelian vector multiplets which, by virtue of the gauge field dualization to a scalar, can be themselves dualized to twisted hypermultiplets.

The previous description breaks down at subvarieties of the Coulomb branch where the residual gauge group is non-abelian. In particular it fails to describe the origin of the Coulomb branch, which flows to a SCFT in the IR. The dualization of a non-abelian vector multiplet is not understood. Instead, a more fruitful exposition takes advantage of special disorder operators, which can be defined directly at the infrared fixed point \cite{Borokhov:2002ib} and which are not polynomial in the microscopic degrees of freedom: they are called 't Hooft monopole operators and are defined by prescribing a Dirac monopole singularity at an insertion point in the Euclidean path integral \cite{'tHooft:1977hy}. Monopole operators are classified by embedding $U(1)\hookrightarrow \cG$, and are labeled by magnetic charges which, by a generalized Dirac quantization \cite{Englert:1976ng}, take value in the weight lattice $\Gamma_{\cG^\vee}$ of the GNO or Langlands dual group $\cG^\vee$ \cite{Goddard:1976qe,Kapustin:2005py}. 
The monopole flux breaks the gauge group $\cG$ to a residual gauge group $H_m$ by the adjoint Higgs mechanism. 
Restricting to gauge invariant monopole operators is achieved by modding out by the Weyl symmetry group, thus restricting $m \in \Gamma_{\cG^\vee}/\cW_\cG$. 

In a three-dimensional $\cN=2$ theory one can define half-BPS monopole operators which sit in chiral multiplets. Crucially, there exists a \emph{unique} BPS monopole operator $V_m$ for each choice of magnetic charge $m$ \cite{Borokhov:2002cg}. If the theory has $\cN=4$ supersymmetry, the $\cN=4$ vector multiplet decomposes into an $\cN=2$ vector multiplet $V$ and a chiral multiplet $\Phi$ in the adjoint representation. To describe the Coulomb branch, $V$ is replaced by monopole operators $V_m$, which now can be dressed by the classical complex scalar $\phi$ inside $\Phi$. This dressing preserves the same supersymmetry of a chiral multiplet \cite{Borokhov:2003yu} if and only if $\phi$ is restricted to $\phi_m$, a constant element of the Lie algebra of the residual gauge group $H_m$ \cite{Cremonesi:2013lqa}. The monopole operators which parametrise the Coulomb branch of an $\cN=4$ field theory are thus polynomials of $V_m$ and $\phi_m$, which are made gauge invariant by averaging over the action of the Weyl 
group \cite{Cremonesi:2013lqa}.  

In this paper the gauge group $\cG$ will mostly be be a product of $U(N_i)$ unitary groups, which are self-dual. For $U(N)$ monopole operators $V_{\vec m}$, with magnetic charge  $\vec m=\diag(m_1,...,m_N)$, the weight lattice of the dual group is given by $\Gamma_{U(N)}=\bZ^N = \left\{ m_i \in \BZ, i=1,..,N \right\}$. Modding out by the Weyl group $S_N$ restricts the lattice to the Weyl chamber $\Gamma_{U(N)}/S_N= \left\{\vec m \in \bZ^N| m_1 \geq m_2 \geq ... \geq m_N\right\}$.

For $U(N)$ gauge groups, which are not simply connected, the center $\CZ (\cG^\vee) = U(1)$ engenders a topological $U(1)_J$ symmetry group. Classically, monopole operators are only charged under this symmetry. To each such $U(N_i)$ gauge group, we associate a fugacity $z_i$ for the topological $U(1)_{J_i}$ symmetry with conserved current $\ast \Tr F_i$, where $F_i$ is the field strength of the $i$-th gauge group. 
Other charges are acquired quantum-mechanically: in particular, monopole operators become charged under the 
Cartan $U(1)_C$ of the $SU(2)_C$ R-symmetry acting on the Coulomb branch. For a Lagrangian $\cN=4$ gauge theory, this charge is given by the formula 
\bea\label{dimension}
\Delta(\vec m)=-\sum_{{\vec \alpha} \in \Delta_{+}}{\left|\vec \alpha(\vec m) \right|} + \frac{1}{2} \sum_{i=1}^{n}{\sum_{{\vec \rho}_i \in \cR_{i}}{\left|{\vec \rho}_{i}(\vec m) \right|}} ~,
\eea
where the first contribution, arising from vector multiplets, is a sum over the positive roots of the gauge group, while the second contribution is a sum over the weights of the gauge group representations of the hypermultiplets. The fugacity for this $R$-symmetry is called $t^2$ in the following. The dimension formula \eqref{dimension} was conjectured in \cite{Gaiotto:2008ak} based on a weak coupling computation in \cite{Borokhov:2002cg}, and later proven exactly in \cite{Benna:2009xd,Bashkirov:2010kz}.
For the theories that we will be studying, which are good or ugly in the sense of \cite{Gaiotto:2008ak}, \eqref{dimension} is believed to equal the scaling dimension in the IR CFT. 

For gauge theories described by (possibly non-simply laced) Dynkin diagrams, we propose the following prescription for computing the $R$-charge of a monopole operator, generalizing the Lagrangian formula \eqref{dimension}. Each diagram is constructed from two basic building blocks: a node and a line.  A $U(N)$ node, with magnetic charge $\vec m$, contributes to the Coulomb branch Hilbert series as follows:
\bea \label{dimension_vector}
\begin{tikzpicture}
    \tikzset{mycirc/.style = {draw, circle, minimum size=0.5mm}}
		\tikzset{
doublearrow/.style={draw, thin, double distance=4pt, ->},
thirdline/.style={draw, thin, ->}
}
    \node (nd1) [mycirc,label=above: \scriptsize $U(N)$] {};
	\end{tikzpicture}
\qquad  \qquad 
\Delta_{\text{vec}}(\vec m)= - \sum_{1\leq i < j \leq N}  
\left|m_i-m_j \right|~.
\eea
A line connecting the nodes $U(N_1)$ and $U(N_2)$ can be either a single bond ($-$), a double bond ($\Rightarrow$) or a triple bond ($\Rrightarrow$), which we take to be oriented from node $1$ to node $2$. Let us assign magnetic charges $\vec m^{(1)}$ and $\vec m^{(2)}$ to $U(N_1)$ and $U(N_2)$ respectively. We propose that the contribution from a line is:
\bea \label{dimension_hyper_guess}
\begin{tikzpicture}
    \tikzset{mycirc/.style = {draw, circle, minimum size=0.5mm}}
		\tikzset{
doublearrow/.style={draw, thin, double distance=4pt, ->},
thirdline/.style={draw, thin, ->}
}
    \node (nd1) [mycirc,label=above: \scriptsize $U(N_1)$] {};
    \node (nd2) [mycirc, right=of nd1, label=above:\scriptsize $U(N_2)$] {};
    \draw[dashed]   (nd1)--(nd2) ;
	\end{tikzpicture}
\qquad 
\Delta_{\text{hyp}}(\vec m^{(1)},\vec m^{(2)})=\frac{1}{2}\sum_{j=1}^{N_1} \sum_{k=1}^{N_2}{\left|\lambda m^{(1)}_j-m^{(2)}_k \right|}
\eea
where $\lambda=1$ for a single bond, $\lambda=2$ for a double bond and $\lambda=3$ for a triple bond. If $\lambda>1$, \eqref{dimension_hyper_guess} does not arise from matter fields transforming in a genuine representation of $U(N_1)\times U(N_2)$.%
\footnote{Conceivably, this prescription could be derived from a Lagrangian quiver gauge theory associated to an unfolded simply laced quiver, further orbifolded by an outer automorphism group of the quiver. We will not pursue this possibility here. We thank Jan Troost for discussions on this point.} 

We stress that formula \eqref{dimension_hyper_guess} is the crucial ingredient that will allow us to compute the Hilbert series of instanton moduli spaces for any simple Lie group. 
We will successfully test our proposal by comparing with known results and by studying general properties of the Hilbert series that can be extracted from the Coulomb branch formula. 

The dimension formula, given by the sum of the two contributions, \eqref{dimension_vector} for each node and \eqref{dimension_hyper_guess} for each line, makes the quivers associated to the affine Dynkin diagrams (\emph{i.e.} before adding the over-extended node) balanced in the sense of \cite{Gaiotto:2008ak}: each unitary gauge group has an effective number of flavors equal to twice the number of colors.%
\footnote{The effective number of flavors for a gauge group is obtained by adding up the ranks of all the gauge groups connected to it by an edge, appropriately weighted by $\lambda$. For instance, for $F_4$ node $2$ has $3k$ colors and $2k+2(2k)=6k$ effective flavors, while node $3$ has $2k$ colors and $3k+k=4k$ flavors. 
} 

%


Once we have classified gauge invariant chiral operators (classical operators, bare and dressed monopole operators) on the Coulomb branch of non-simply laced quivers by their quantum number $J_i$ and $\Delta$, we enumerate them by means of a generating function that grades them by their charges. The Hilbert series of the Coulomb branch of a $d=3$ $\cN=4$ good or ugly superconformal field theory is then given by  \cite{Cremonesi:2013lqa}
\be\label{ref_HS}
HS(t,\vec{z})=\sum_{\vec{m}\in \Gamma_{\cG^\vee}/\cW_{\cG}} \vec{z}^{\vec{J}(\vec{m})}~ t^{2\Delta(\vec{m})} P_{\cG}(t;\vec{m})~,
\ee
where $\vec{z}^{\vec{J}(\vec{m})}=\prod_i z_i^{J_i(m)}$.
The sum is over GNO magnetic sectors \cite{Goddard:1976qe}, restricted to a Weyl chamber to impose invariance under the gauge group $\cG$. There is one bare monopole operator per magnetic charge sector \cite{Borokhov:2002cg}.  
The factors $\vec{z}^{\vec{J}(\vec{m})}~ t^{2\Delta(\vec{m})}$ account for the topological charges and conformal dimension of bare monopole operators of magnetic charge $\vec{m}$. Finally, the factor $P_{\cG}(t;\vec{m})$ reflects the dressing of a bare monopole operator $V_{\vec{m}}$ by polynomials of the classical adjoint scalar $\phi_m\in \mathfrak{h}_{\vec{m}}$ which are gauge invariant under the residual gauge group $H_{\vec{m}}$ left unbroken by the monopole flux. The contribution of this dressing factor to the Hilbert series is given by the generating function of $H_{\vec{m}}$ Casimir invariants
\bea\label{Casimir} 
P_\cG(t; \vec{m})=\prod_{i=1}^{\mathrm{rk}(\cG)} \frac{1}{1-t^{2d_i(\vec{m})}}
\eea
where $d_i(\vec{m})$ are the degrees of the Casimir invariants of $H_{\vec{m}}$.%
\footnote{\eqref{Casimir} assumes that the ring of Casimir invariants is freely generated, as is the case for semisimple Lie groups. The assumption could fail if the gauge group contains extra discrete factors, in which case \eqref{Casimir} is to be replaced by the appropriate Molien formula. We will not encounter this subtlety in this paper.} We refer the readers to Appendix A of \cite{Cremonesi:2013lqa} for more details on these classical dressing factors.

In the next sections we will apply formula \eqref{ref_HS} to the non-simply laced quivers discussed in section ~\ref{sec:quiversAndBranes} and compute exactly the Hilbert series of the corresponding three instanton moduli spaces. To make contact with moduli spaces of $G$-instantons, we first need to specify how the fugacities $\vec{z}$ of the topological symmetry are related to the fugacities $x$ and $\vec{u}$ of the global $SU(2)_x \times G_{\vec{u}}$ symmetry acting on $G$-instantons.

\subsection{Refinement}

Consider a generalized quiver gauge theory corresponding to an over-extended affine Dynkin diagram from Tables \ref{tab:instantonBraneConstCoulombBranch1} and \ref{tab:instantonBraneConstCoulombBranch2}. We label the nodes as follows: $i=1,\dots,r=\mathrm{rk}(G)$ for the nodes of the Dynkin diagram of the finite Lie algebra $\mathrm{Lie}(G)$, $i=0$ for the affine node corresponding to the null root, and $i=-1$ for the over-extended node attached to the $i=0$ node. The ranks $N_i$ of the associated unitary groups are given by $N_{-1}=1$ for the over-extended node and by $N_i=k a_i^\vee$, $i=0,\dots, r$, for the nodes of the affine Dynkin diagram. Each unitary gauge group has a topological symmetry $U(1)_{J_i}$ with fugacity $z_i$. 

When all the nodes are treated as gauge groups, an overall diagonal $U(1)$ is decoupled and needs to be factored out. This decoupled $U(1)$ corresponds to the shift symmetry 
\be\label{decoupled_U(1)}
m_{-1} \to m_{-1} + c~, \qquad m_i \to m_i +c~\frac{a_i}{a_i^\vee} ~\unit_{k a_i^\vee} \qquad (i=0,\dots r)~, \qquad c \in \bZ
\ee
in the dimension formula, where $\unit_n$ denotes the $n\times n$ unit matrix, $a_i$ are the Coxeter labels and $a_i^\vee$ are the dual Coxeter labels of the untwisted affine algebra (in particular $a_0=a_0^\vee=1$). 
Note that for the untwisted affine algebras the ratio $a_i/a_i^\vee$ is an integer. The decoupled $U(1)$ is factored out by fixing the shift symmetry \eqref{decoupled_U(1)}, multiplying the Coulomb branch Hilbert series by its inverse classical factor $(1-t^2)$, and setting to $1$ the fugacity of the associated topological symmetry: 
\be\label{fugacity_constraint}
z_{-1} \bigg(\prod_{i=0}^r  z_i^{a_i} \bigg)^k=1~.
\ee

The constraint \eqref{fugacity_constraint} on the fugacities ensures that the shift \eqref{decoupled_U(1)} does not affect the Hilbert series and determines $z_{-1}$ in terms of the remaining $r+1$ fugacities $z_i$, $i=0,\dots,r$, associated to the nodes of the untwisted affine Dynkin diagram. The fugacities $z_i$, $i=1,\dots,r$, associated to the nodes of the Dynkin diagram of $\mathrm{Lie}(G)$ are simple root fugacities for the global symmetry $G$, therefore in \eqref{nonfactorisableModSpace} we can identify 
\be\label{y_equal_z}
u_i=z_i~, \qquad i=1,\dots,r~.
\ee 

The fugacity $x$ for the $SU(2)$ rotational symmetry is determined by identifying the two unique monopole operators of dimension $\Delta=\frac{1}{2}$, which generate the $\bC^2$ moduli space of the center of mass of the instantons. 
The tower of monopole operators obtained by rescaling these magnetic fluxes by an integer then reconstructs the prefactor in \eqref{nonfactorisableModSpace}. Let us focus on a monopole operator which generates a $\bC$ subspace of the $\bC^2$ moduli space of the center of mass, and assign to it weight $t x$ in the Hilbert series for definiteness.%
\footnote{The monopole operator with weight weight $t x^{-1}$ is obtained by flipping sign to the magnetic flux and acting with the Weyl group to bring the resulting flux to the positive Weyl chamber.} 
Up to the shift \eqref{decoupled_U(1)}, the magnetic charge of this monopole operator (written in matrix notation) can be taken to be%
\footnote{We use the shorthand notation $(r^s)=(\underbrace{
r,\cdots,r}_\text{$s$ times})$.} 
\be\label{monopole_generating_C}
m_{-1}=0~, \qquad 
m_i= \diag(1,0^{k-1}) \otimes \frac{a_i}{a_i^\vee}~\unit_{a_i^\vee}
~, \quad i=0,\dots,r~.
\ee
It is straightforward to see that the monopole operator with magnetic charge \eqref{monopole_generating_C} has dimension $\Delta=\frac{1}{2}$: because the contributions to $\Delta$ coming from the untwisted affine Dynkin diagram cancel out, while the contribution of the edge connecting the extended node to the over-extended node is $\frac{1}{2}$. 
From the topological charge of the monopole operator of magnetic charge \eqref{monopole_generating_C} we read off the fugacity for the $SU(2)_x$ rotational symmetry,
\be\label{x_definition}
x =\prod_{i=0}^r z_i^{a_i} = z_0 \prod_{i=1}^r u_i^{a_i} ~.
\ee
In the last equality we have used $a_0=1$ and the identification \eqref{y_equal_z}. \eqref{x_definition} can be used to express $z_0$ in terms of $x$ and $\vec{u}$. The constraint \eqref{fugacity_constraint} from the removal of the decoupled $U(1)$ then determines $z_{-1}$ as 
\be\label{fugacity_constraint_2}
z_{-1}= x^{-k}~.
\ee

\section{$k$ $G_2$ instantons}\label{sec:k-G2}
The theory whose Coulomb branch is the moduli space of $k$ $G_2$ instantons on $\BC^2$ is described by the quiver diagram
%
%
\bea \label{QuivG2}
{\blue \node{}{1}}-\Node{}{k}-\node{}{2k}\Rrightarrow \node{}{k} \ \tikz[na]\node(quivG2){};
\eea
where each number denotes the rank of each unitary gauge group and an overall $U(1)$ symmetry is factored out.

The dimension formula for $k$ $G_2$ instantons can be extracted from this quiver using the prescription of Section \ref{sec:Hilbert}:
\be \label{G2Delta}
\begin{split}
\Delta_{k,G_2}(\vec m, \vec n, \vec s) &= \sum_{i=1}^k |m_i| + \sum_{i=1}^k \sum_{j=1}^{2k} |m_i - n_j|+ \sum_{j=1}^{2k} \sum_{\ell=1}^{k} |3n_j - s_\ell| \\
& -2 \left( \sum_{1 \leq i<i' \leq k} |m_i-m_{i'}| +\sum_{1 \leq j<j' \leq 2k} |n_j-n_{j'}|+\sum_{1 \leq \ell<\ell' \leq k} |s_\ell-s_{\ell'}| \right), 
\end{split}
\ee
where $\vec m=(m_1,...,m_k)$, $\vec n=(n_1,...,n_{2k})$ and $\vec s=(s_1,...,s_k)$. Note the factor of 3 in front of $n_j$ for the triply laced bifundamental contribution. Here we have gauge fixed the decoupled $U(1)$ by setting the monopole flux of the over-extended node (indicated in blue) to zero.

The Hilbert series for the moduli space of $k$ $G_2$ instantons can thus be computed as follows:
\be \label{G2HS}
\begin{split}
g_{k, G_2} (t; \vec z) &= \sum_{m_1 \geq  \cdots \geq m_k > -\infty}~ \sum_{n_1 \geq  \cdots \geq n_{2k} > -\infty} ~ \sum_{s_1 \geq  \cdots \geq s_{k} > -\infty} t^{\Delta_{k,G_2}(\vec m, \vec n, \vec s)}  \\
& \quad P_{U(k)} (t; \vec m) P_{U(2k)} (t; \vec n) P_{U(k)} (t; \vec s)  \times z_0^{\sum_{i=1}^k m_i} z_1^{\sum_{j=1}^{2k} n_j} z_2^{\sum_{\ell=1}^{k} s_\ell}~,
\end{split}
\ee
where the fugacities $\vec z$ are associated to the topological symmetry. 

For $k=1$, the result of \eqref{G2HS} can be written as
\bea
g_{1, G_2}(t; \vec z) = \frac{1}{(1- tx)(1-t x^{-1})} \sum_{p=0}^\infty \chi^{G_2}_{[p,0]}(u_1, u_2)  t^{2p}~,
\eea
where $[1,0]$ is the adjoint representation of $G_2$ and
\bea
x= z_0 z_1^2 z_2^3~,  \qquad u_1 = z_1, \qquad  u_2 = z_2~.
\eea
This agrees with (5.46) of \cite{Benvenuti:2010pq}.

It is worth mentioning that, for $k\geq 2$, the Hilbert series \eref{G2HS} can alternatively be computed using the Hall-Littlewood formula and the gluing technique discussed in \cite{Cremonesi:2014kwa, Cremonesi:2014vla}. Indeed quiver \eref{QuivG2} can be constructed by gluing the following two basic building blocks
\bea \label{twobuildingblocksG2}
T_{(k, k-1, 1)}(SU(2k)): \; (1)-(k)-[2k]~, \qquad 
T_{(k, k)}(SU(2k)): \; [2k]-(k)~,
\eea
once the edge $[2k]-(k)$ in the second building block is converted to $[2k]\Rrightarrow(k)$ by tripling the value of the background magnetic charges in the Coulomb branch Hilbert series of $T_{(k,k)}(SU(2k))$.  The two building blocks are glued by gauging the common flavor symmetry $U(2k)/U(1)$. 
The final expression of the Hilbert series in question is given by 
\be
\begin{split}
g_{k, G_2} (t; \vec a, \vec b) &= \sum_{n_1 \geq n_2\geq  \ldots \geq n_{2k-1} \geq n_{2k} = 0} t^{-2\delta_{U(2k)}(\vec n)} (1-t^2) P_{U(2k)} (t; n_1, \ldots, n_{2k}) \times \\
&  \qquad H[ T_{(k, k-1, 1)}(SU(2k))] (t; a_1, a_2, a_3; n_1, \ldots, n_{2k}) \times  \\
&  \qquad H[ T_{(k, k)}(SU(2k))] (t; b_1, b_2, b_3; 3n_1, \ldots, 3n_{2k})~.
\end{split}
\ee
The Hall-Littlewood formulae for the Coulomb branch HS of \eref{twobuildingblocksG2} are given by
\bea
\begin{split}
&H[ T_{(k, k-1, 1)}(SU(2k))] (t;a_1, a_2, a_3; \vec n)  \\
& \qquad = t^{\delta_{U(2k)}(\vec n)} (1-t^2)^{2k} K_{(k,k-1,1)}(t; a_1, a_2,a_3) \Psi^{\vec n}_{U(2k)} (\vec v_{(k,k-1,1)}; t)~, 
\end{split} \\
\begin{split}
&H[ T_{(k, k)}(SU(2k))] (t; b_1, b_2; \vec n) \\
& \qquad = t^{\delta_{U(2k)}(\vec n)} (1-t^2)^{2k} K_{(k,k)}(t; b_1, b_2) \Psi^{\vec n}_{U(2k)} (\vec v_{(k,k)}; t)~, \label{HSTkk}
\end{split}
\eea
where the Hall-Littlewood polynomial is defined as 
\bea
\Psi^{\vec n}_{U(N)} (x_1,\dots,x_N;t)=\sum_{\sigma \in S_N}
x_{\sigma(1)}^{n_1} \dots x_{\sigma(N)}^{n_N}
\prod_{1 \leq i<j \leq N}   \frac{  1-t x_{\sigma(i)}^{-1} x_{\sigma(j)} } {1-x_{\sigma(i)}^{-1} x_{\sigma(j)}}~,
\eea
and the parameters and prefactors are given by%
\footnote{The plethystic exponential ($\PE$) of a multi-variate function $f(x_1, \ldots, x_n)$ is defined as
\bea
\PE \left[f(x_1, \ldots, x_n) \right] = \exp \left(\sum_{p=1}^\infty \frac{1}{p} f(x_1^p, \ldots, x_n^p) \right)~. \nn
\eea
}
\bea
\delta_{U(2k)}(\vec n) &= \sum_{1\leq i < j \leq 2k} (n_i-n_j)~,  \\
\vec v_{(k,k-1,1)} &=   \Big(t^{k-1} a_1, t^{k-3} a_1, \ldots, t^{-(k-3)} a_1 , t^{-(k-1)} a_1,  \nn \\
& \qquad t^{k-3} a_2, t^{k-5} a_2, \ldots, t^{-(k-5)} a_2, t^{-(k-3)} a_2, a_3  \Big) ~, \nn \\
\vec v_{(k,k)} &= \Big(t^{k-1} b_1, t^{k-3} b_1, \ldots, t^{-(k-3)} b_1 , t^{-(k-1)} b_1,  \nn \\
& \qquad t^{k-1} b_2, t^{k-3} b_2, \ldots, t^{-(k-3)} b_2, t^{-(k-1)} b_2  \Big) ~, \nn \\
K_{(k,k-1,1)}(t; \vec a) &= \PE \Bigg[(t^2+t^{2k}) + 2\sum_{m=1}^{k-1} t^{2m} +(a_2 a_3^{-1}+a_2^{-1} a_3) t^{k} +(a_1 a_3^{-1}+a_1^{-1} a_3) t^{k+1}  \nn \\
& \hspace{4cm} + (2+a_1 a_2^{-1}+a_2 a_1^{-1})\sum_{m=1}^{k} t^{2m-1} \Bigg]~, \nn \\
K_{(k,k)}(t; \vec b) &= \PE \left[  \left(2+ b_1 b_2^{-1} +b_1^{-1} b_2  \right) \sum_{m=1}^{k} t^{2m} \right]~. \nn
\eea
The fugacities can be set as follows:
\bea
a_1^k a_2^{k-1} a_3 =1~,  \qquad b_1^k b_2^k =1~.
\eea

The relations between the fugacities $\vec a$ and $\vec b$ to the topological fugacity of each node in quiver \eref{QuivG2} are given by (see (3.13) of \cite{Cremonesi:2014kwa})
\bea
z_{-1}= a_3 a_2^{-1}, \quad z_0= a_2 a_1^{-1}, \quad z_1 = a_1 b_1^3, \quad z_2 = b_2 b_1^{-1}~,
\eea 
and by factoring out the overall $U(1)$ we have the following condition (cf. (3.3) of \cite{Cremonesi:2014vla}):
\bea
z_{-1} (z_0 z_1^2 z_2^3)^k =1~.
\eea

From \eref{y_equal_z} and \eref{fugacity_constraint_2}, we find that the relations between $\vec a, \vec b$ and the fugacities $x$ associated with $SU(2)$ and $u_1, u_2$ associated with $G_2$ are
\bea
& x= z_0 z_1^2 z_2^3 = a_1 a_2 (b_1 b_2)^3~, \nn \\
& z_1 = a_1 b_1^3~, \quad z_2 = b_2 b_1^{-1}~.
\eea

For $k=2$ we recover the Hilbert series (9.3) and (9.5)%
\footnote{There is a typo in Eq. (9.5) of \cite{Hanany:2012dm}: the power of  $(1+t+t^2)$ in the denominator should be $7$.} of \cite{Hanany:2012dm}.  

For $k=3$ let us report only the result with $z_i$ being set to unity; the unrefined Hilbert series of the reduced three $G_2$ instanton moduli space is \be
\begin{split} \label{3G2HSur}
\tilde{g}_{3, G_2}(t) &= \frac{1-t}{(1-t^2)^{7} (1-t^3)^9 (1-t^4)^7} 
\Big(1+t+11 t^2+34 t^3+124 t^4 +352 t^5\\
& +1055 t^6+2657 t^7+6584 t^8+14635 t^9+31194 t^{10}+61229 t^{11}  \\
& +114367 t^{12}+198932 t^{13}+329172 t^{14}+511194 t^{15}+755093 t^{16} \\
&  +1051845 t^{17}+1394817 t^{18}+1749632 t^{19}+2091341 t^{20}+2368619 t^{21} \\
& +2557449 t^{22}+2619060 t^{23}+2557449 t^{24} + \text{palindrome up to $t^{46}$} \Big) ~.
\end{split}
\ee

\section{$k$ $B_N$ instantons} \label{sec:k-SO(2N+1)}
The theory whose Coulomb branch is the moduli space of $k$ $SO(2N+1)$ instantons on $\BC^2$ is described by the quiver diagram
%
%
\bea \label{quivB}
{\blue \node{}{1}}-\Node{}{k}-\node{{\ver{}{k}}}{2k}-\underbrace{\node{}{2k}-\cdots-\node{}{2k}}_{N-3~\text{nodes}}\Rightarrow\node{}{k}  \ \tikz[na]\node(quiv1){};
\eea
where each number denotes the rank of a unitary gauge group and the decoupled overall $U(1)$ symmetry is removed.  
For $k=2$ we recover the results given in Section 5 of \cite{Hanany:2012dm}.


The unrefined Hilbert series of the reduced $3$ $SO(7)$ instanton moduli space is 
\bea
\tilde{g}_{3, SO(7)}(t) &= \frac{(1-t)^2}{(1-t^2)^{9} \left(1-t^3\right)^{12} \left(1-t^4\right)^9} 
\Big(1+2 t+18 t^2+68 t^3+292 t^4+1024 t^5 \nn \\
& +3565 t^6+11012 t^7+32587 t^8+88764 t^9+229405 t^{10}+554642 t^{11} \nn \\
& +1271439 t^{12}+2749154 t^{13}+5648717 t^{14}+11006976 t^{15}+20431264 t^{16} \nn\\
& +36104898 t^{17}+60918929 t^{18}+98135686 t^{19}+151245678 t^{20}+223030062 t^{21} \nn \\
& +315153966 t^{22}+426792414 t^{23}+554536028 t^{24}+691345362 t^{25} \nn \\
& +827700194 t^{26}+951603050 t^{27}+1051256831 t^{28}+1115766454 t^{29} \nn \\
& +1138239548 t^{30}+1115766454 t^{31}+ \text{palindrome up to $t^{60}$} \Big)~.
\eea

\section{$k$ $C_N$ instantons} \label{sec:k-USp(2N)}

The theory whose Coulomb branch is the moduli space of $k$ $USp(2N)$ instantons on $\BC^2$ is described by the quiver diagram
%
\bea \label{quivC}
{\blue \node{}{1}}-\Node{}{k}\Rightarrow \underbrace{\node{}{k}-\cdots-\node{}{k}}_{N-1~\text{nodes}}\Leftarrow\node{}{k}  \ \tikz[na]\node(quivC){};
\eea
where each number denotes the rank of a unitary gauge group and an overall $U(1)$ symmetry decouples.  For $k=2$ we recover the results given in Section 4.2 of \cite{Hanany:2012dm}.  Below we present the unrefined Hilbert series for $3$ instantons and small values of $N$.


The unrefined Hilbert series of the reduced $3$ $USp(4)$ instanton moduli space is 
\be
\begin{split}
\tilde{g}_{3, USp(4)}(t) &= \frac{1}{(1-t^2)^{5} (1-t^3)^{6} (1-t^4)^5} 
\Big( 1+8 t^2+18 t^3+61 t^4+142 t^5 \\
&  +388 t^6+792 t^7+1691 t^8+2996 t^9+5255 t^{10}+7994 t^{11}+11713 t^{12}  \\
&  +15134 t^{13}+18773 t^{14}+20796 t^{15}+21980 t^{16}+20796 t^{17}+18773 t^{18}  \\
&  + \text{palindrome up to $t^{32}$} \Big)~.
\end{split}
\ee
The unrefined Hilbert series of the reduced $3$ $USp(6)$ instanton moduli space is 
\be
\begin{split}
\tilde{g}_{3, USp(6)}(t) &= \frac{1}{(1-t^2)^7 (1-t^3)^8 (1-t^4)^7}
\Big(1+17 t^2+38 t^3+209 t^4+644 t^5  \\
&  +2260 t^6+6382 t^7+17808 t^8+43106 t^9+99660 t^{10}+206484 t^{11} \\
&  +404244 t^{12} +724452 t^{13}+1224332 t^{14}+1917162 t^{15}+2834175 t^{16} \\
& +3909874 t^{17}+5102043 t^{18} +6239722 t^{19}+7227435 t^{20}+7864776 t^{21} \\
& +8110736 t^{22}+7864776 t^{23}+ \text{palindrome up to $t^{44}$} \Big)~.
\end{split}
\ee
The unrefined Hilbert series of the reduced $3$ $USp(8)$ instanton moduli space is 
\bea
\tilde{g}_{3, USp(8)}(t) &=\frac{1}{(1-t^2)^9 (1-t^3)^{10} (1-t^4)^9} 
\Big(1+30 t^2+66 t^3+564 t^4+1978 t^5 \nn \\
& +8986 t^6+31320 t^7+108588 t^8+327552 t^9+938028 t^{10}+2428438 t^{11} \nn \\
& +5923950 t^{12}+13333518 t^{13}+28288029 t^{14}+56057448 t^{15}+105000098 t^{16}\nn \\
& +185111036 t^{17}+309423948 t^{18}+489269266 t^{19}+735494922 t^{20} \nn \\
& +1049537386 t^{21}+1426754090 t^{22}+1845578580 t^{23}+2277688217 t^{24} \nn \\
& +2678999920 t^{25}+3009187465 t^{26}+3224258916 t^{27}+3300770520 t^{28} \nn \\
& +3224258916 t^{29}+3009187465 t^{30} + \text{palindrome up to $t^{56}$} \Big)~.
\eea
The unrefined Hilbert series of the reduced $3$ $USp(10)$ instanton moduli space is 
\bea
\tilde{g}_{3, USp(10)}(t) &=\frac{(1-t)^2}{(1-t^2)^{13} (1-t^3)^{12} (1-t^4)^{11}} \Big(1+2 t+48 t^2+196 t^3+1533 t^4+7458 t^5 \nn \\
& +39083 t^6+173746 t^7+729193 t^8+2753342 t^9+9659061 t^{10}+31142740 t^{11} \nn \\
& +93620178 t^{12}+262065600 t^{13}+688287079 t^{14}+1698315214 t^{15}+3955023058 t^{16} \nn \\
& +8708306700 t^{17}+18185341012 t^{18}+36076921166 t^{19}+68144856266 t^{20} \nn \\
& +122727426896 t^{21}+211098608616 t^{22}+347187234006 t^{23}+546680541199 t^{24} \nn \\
& +824886510488 t^{25}+1193911094540 t^{26}+1658736457996 t^{27}+2213773962229 t^{28} \nn \\
& +2839692757258 t^{29}+3502903178369 t^{30}+4156849878890 t^{31}+4747242880506 t^{32} \nn \\
& +5218604879584 t^{33}+5523278387053 t^{34}+5628609146268 t^{35}+5523278387053 t^{36} \nn \\
& + \text{palindrome up to $t^{70}$} \Big)~.
\eea

For higher number of instantons, the Hilbert series can be computed more easily from the Higgs branch of the ADHM quiver.  We demonstrate this computation in Appendix \ref{sec:USp2NHiggs}. Let us report here the unrefined Hilbert series (\ie~ $x=1$ and $z_i=1$ for all $i$) for $k=5$ and small values of $N$:
\bea
{\tilde g}_{5, USp(2)}(t) &= \frac{1}{(1-t^2)^{4} (1-t^3)^{4}(1-t^4)^3 (1-t^5)^4 (1-t^6)^3 } \times \nn \\
& \Big( 1+2 t^2+6 t^3+14 t^4+26 t^5+59 t^6+108 t^7+216 t^8+382 t^9+669 t^{10}+1090 t^{11} \nn \\
& +1788 t^{12}+2718 t^{13}+4080 t^{14}+5844 t^{15}+8166 t^{16}+10902 t^{17}+14271 t^{18}\nn \\
& +17886 t^{19}+21899 t^{20}+25824 t^{21}+29701 t^{22}+32898 t^{23}+35621 t^{24}+37152 t^{25} \nn \\
& +37792 t^{26}+37152 t^{27}+ \text{palindrome up to $t^{52}$} \Big)~. \nn \\
{\tilde g}_{5, USp(4)}(t) &= \frac{1}{(1-t^2)^5 (1-t^3)^6 (1-t^4)^6 (1-t^5)^6 (1-t^6)^5 } \times \nn \\
& \Big ( 1+8 t^2+18 t^3+65 t^4+184 t^5+568 t^6+1486 t^7+4068 t^8+10202 t^9+25294 t^{10} \nn \\
& +59530 t^{11}+136840 t^{12}+301276 t^{13}+645420 t^{14}+1332274 t^{15}+2669897 t^{16} \nn \\
& +5173382 t^{17}+9731196 t^{18}+17732334 t^{19}+31384129 t^{20} +53895904 t^{21}\nn \\
& +89958111 t^{22}+145882550 t^{23}+230128561 t^{24}+353099760 t^{25}+527468664 t^{26} \nn \\
& +767161840 t^{27}+1087152304 t^{28}+1501274126 t^{29}+2021417792 t^{30}+2654217372 t^{31} \nn \\
& +3400290035 t^{32}+4250584996 t^{33}+5186895160 t^{34}+6179265798 t^{35}+7189118462 t^{36} \nn \\
& +8168673774 t^{37}+9067212695 t^{38}+9832235886 t^{39}+10417596422 t^{40}+10784743772 t^{41} \nn \\
& +10910252456 t^{42}+10784743772 t^{43}+ \text{palindrome up to $t^{84}$} \Big)~.
\eea

\section{$k$ $F_4$ instantons} \label{sec:k-F4}

The theory whose Coulomb branch is the moduli space of $k$ $F_4$ instantons on $\BC^2$ is described by the quiver diagram
\bea \label{QuivF4}
{\blue \node{}{1}}- \Node{}{k}-\node{}{2k}-\node{}{3k}\Rightarrow\node{}{2k}-\node{}{k} \quad\tikz[na]\node(quivF4){};
\eea
where each number denotes the rank of a unitary gauge group and an overall $U(1)$ symmetry is factored out.  

The Hilbert series of $k$ $F_4$ instantons can be computed using the monopole formula given by \eref{ref_HS}.
For $k\geq 2$, \eref{ref_HS} is more easily calculated using the gluing technique discussed in \cite{Cremonesi:2014vla}. Indeed quiver \eref{QuivF4} can be constructed from the building blocks
\be \label{buildingblocksF4}
\begin{split}
T_{(k, k, k-1, 1)}(SU(3k)):&  \quad  (1)-(k)-(2k)-[3k]~, \\
T_{(k,k,k)}(SU(3k)):& \quad  [3k]-(2k)-(k)~,
\end{split}
\ee
once the edge $[3k]-(2k)$ in the second building block is converted to $[3k]\Rightarrow(2k)$ by doubling the value of the background magnetic charges in the Coulomb branch Hilbert series of $T_{(k,k,k)}(SU(3k))$. The two building blocks are glued by gauging the common flavor symmetry $U(3k)/U(1)$. 

The final expression of the Hilbert series in question is given by 
\be
\begin{split}
g_{k, F_4} (t; \vec a, \vec b) 
&= \sum_{m_1 \geq m_2\geq  \ldots \geq m_{3k}=0} t^{-2\delta_{U(3k)} (\vec m)} (1-t^2) P_{U(3k)} (t; m_1, \ldots, m_{3k}) \times  \\
&  \qquad H[ T_{(k, k, k-1, 1)}(SU(3k))] (t; a_1, a_2, a_3,a_4; m_1, \ldots, m_{3k}) \times   \\
& \qquad   H[ T_{(k, k,k)}(SU(3k))] (t; b_1, b_2,b_3; 2m_1, \ldots, 2m_{3k})~.
\end{split}
\ee
The Coulomb branch Hilbert series of $T_{(k, k, k-1, 1)}(SU(3k))$ is given by
\be
\begin{split}
&H[ T_{(k, k, k-1, 1)}(SU(2k))] (t;a_1, a_2, a_3,a_4; \vec n) \\
& \qquad = t^{\delta_{U(3k)} (\vec n)} (1-t^2)^{3k} K_{(k,k,k-1,1)}(t; a_1, a_2,a_3,a_4) \Psi^{\vec n}_{U(3k)} (\vec v_{(k,k,k-1,1)}; t)~,
\end{split}
\ee
with 
\bea
\delta_{U(3k)} (\vec n) &= \sum_{1\leq i < j \leq 3k} (n_i-n_j)~,  \\
\begin{split}
\vec v_{(k, k,k-1,1)} &=   \Big(t^{k-1} a_1, t^{k-3} a_1, \ldots, t^{-(k-3)} a_1 , t^{-(k-1)} a_1,   \\
&  \hspace{1cm}  t^{k-1} a_2, t^{k-3} a_2, \ldots, t^{-(k-3)} a_2, t^{-(k-1)} a_2  ~,  \\
&  \hspace{1cm}  t^{k-2} a_3, t^{k-4} a_3, \ldots, t^{-(k-4)} a_3, t^{-(k-2)} a_3, a_4  \Big) ~,  
\end{split}
\\
\begin{split}
K_{(k,k,k-1,1)}(t; \vec a) &= \PE \Bigg[(t^2+t^{2k}) + \sum_{m=1}^{k-1} t^{2m} +(a_3 a_4^{-1}+a_4^{-1} a_3) t^{k}  \\
& \hspace{1cm} +(a_1 a_4^{-1}+a_1^{-1} a_4+a_2 a_4^{-1}+a_2^{-1} a_4) t^{k+1}  \\
& \hspace{1cm} +(a_1 a_3^{-1}+a_1^{-1} a_3+a_2 a_3^{-1}+a_2^{-1} a_3) \sum_{m=1}^k t^{2m-1} \\
& \hspace{1cm}  + (2+a_1 a_2^{-1}+a_2 a_1^{-1})\sum_{m=1}^{k} t^{2m} \Bigg]~.
\end{split}
\eea
On the other hand, the Coulomb branch Hilbert series of $T_{(k, k, k)}(SU(3k))$ is
\be
\begin{split}
&H[ T_{(k, k, k)}(SU(2k))] (t;b_1, b_2, b_3; \vec n)  \\
& \qquad = t^{\delta_{U(3k)} (\vec n)} (1-t^2)^{3k} K_{(k,k,k)}(t; b_1, b_2, b_3) \Psi^{\vec n}_{U(3k)} (\vec v_{(k,k,k)}; t)~,
\end{split}
\ee
with
\bea
\begin{split}
\vec v_{(k, k,k)} &=  \Big(t^{k-1} b_1, t^{k-3} b_1, \ldots, t^{-(k-3)} b_1 , t^{-(k-1)} b_1,  \\
&  \hspace{1cm}  t^{k-1} b_2, t^{k-3} b_2, \ldots, t^{-(k-3)} b_2, t^{-(k-1)} b_2  ~, \\
&  \hspace{1cm}  t^{k-1} b_3, t^{k-3} b_3, \ldots, t^{-(k-3)} b_3, t^{-(k-1)} b_3  \Big) ~,
\end{split}  \\
K_{(k,k,k)}(t; \vec b) &= \PE \left[  \left(\sum_{1\leq i,j\leq 3} b_i b_j^{-1} \right) \sum_{m=1}^{k} t^{2m} \right]~. 
\eea
The fugacities can be set as follows:
\bea
a_1^k a_2^{k} a_3^{k-1} a_4 =1~,  \qquad b_1^k b_2^k b_3^k =1~.
\eea

The relations between the fugacities $\vec a$ and $\vec b$ to the topological fugacity of each node in quiver \eref{QuivF4} are given by (see (3.13) of \cite{Cremonesi:2014kwa}) 
\be
\begin{split}
&z_{-1}= a_4 a_3^{-1}, \quad z_0= a_3 a_2^{-1}, \quad z_1 = a_2 a_1^{-1}, \\
& z_2 = a_1 b_1^2, \quad z_3 = b_2 b_1^{-1}, \quad z_4 = b_3 b_2^{-1}~, 
\end{split}
\ee 
and by factoring out the overall $U(1)$ we have the following condition (cf. (3.3) of \cite{Cremonesi:2014vla}): 
\bea
z_{-1} (z_0 z_1^2 z_2^3 z_3^4 z_4^2)^k = 1~.
\eea

From \eref{y_equal_z} and \eref{fugacity_constraint_2}, we find that the relations between $\vec a, \vec b$ and the fugacities $x$ associated with $SU(2)$ and $u_1, u_2, u_3,u_4$ associated with $F_4$ are
\be
\begin{split}
& x= z_0 z_1^2 z_2^3 z_3^4 z_4^2 = a_1 a_2 a_3 (b_1 b_2 b_3)^2~, \quad  \\
& u_1 = a_2 a_1^{-1}~, \quad u_2 = a_1 b_1^2~, \quad u_3= b_2 b_1^{-1}~, \quad u_4 = b_3 b_2^{-1}~.
\end{split}
\ee

For $k=2$ we recover the results given in (10.2) and (10.4) of \cite{Hanany:2012dm}.


\section{The moduli space of instantons as an algebraic variety} \label{sec:modSpaceAlgebraicVariety}
\subsection{One instanton}
The reduced moduli space of one $G$ instanton is the orbit of the highest root vector in the complexification of the Lie algebra of $G$ \cite{kronheimer1990, Br, KS1}, also known as minimal nilpotent orbit.  The space of holomorphic functions on such a reduced moduli space was studied in \cite{Benvenuti:2010pq}.%
\footnote{See \cite{VinbergPopov, Garfinkle} for a mathematical perspective on this type of varieties, independent of instantons.} 
The Hilbert series can be obtained as
\bea
H(t, \vec u) = \sum_{p=0}^\infty \chi^{G}_{p \cdot {\bf Adj}} (\vec u)  t^{2p}~,  
\eea
where $p \cdot {\bf Adj}$ denotes the irreducible representation of $G$ whose highest weight is $p$ times that of the adjoint representation.  The plethystic logarithm\footnote{The plethystic logarithm of a multi-variate function $f(x_1, \ldots, x_n)$ such that $f(0, \ldots, 0)=1$ is 
\bea
\PL[f(x_1, \ldots, x_n)] =  \sum_{k=1}^\infty \frac{1}{k} \mu(k) \log f(x_1^k, \ldots, x_n^k) ~. \nn
\eea} of this Hilbert series reads
\bea
\PL \left[H(t, \vec u)  \right] =   \chi^{G}_{{\bf Adj}} (\vec u) t^2 - \left(\chi^G_{{\rm Sym}^2 {\bf Adj}}(\vec u) - \chi^G_{{2 \cdot\bf Adj}}(\vec u) \right) t^4 + \ldots~.
\eea
The meaning of the plethystic logarithm is as follows.

The generator $M$ of the reduced moduli space is of order 2 and transforms in the adjoint representation of $G$.  There are relations at order 4 transforming in the representation ${\rm Sym}^2 {\bf Adj} - 2\cdot {\bf Adj}$, where the minus sign means that the irreducible representation $2\cdot {\bf Adj}$ is removed from the decomposition of ${\rm Sym}^2 {\bf Adj}$. These are known as the {\it Joseph relations} \cite{joseph1976minimal} (see also \cite{Gaiotto:2008nz}).  

For the case of $G=SU(N)$, ${\rm Sym}^2 {\bf Adj}$ decomposes as
\bea
{\rm Sym}^2 {\bf Adj} &= {\rm Sym}^2 [1,0,\ldots,0,1] \nn \\
&= [2,0,\ldots,0,2]+[1,0,\ldots,0,1]+[0,\ldots,0]+[0,1,0,\ldots,0,1,0]~.
\eea
Thus,
\bea
{\rm Sym}^2 {\bf Adj} - 2\cdot {\bf Adj} = [1,0,\ldots,0,1]+[0,\ldots,0]+[0,1,0,\ldots,0,1,0]~.
\eea
In this case, the generator $M$ of the reduced moduli space is an $N \times N$ traceless matrix, and the Joseph relations can be explicitly written as
\bea
M^{a_1}_{~a_2} M^{a_2}_{~a_3} =(M^2)^{a_1}_{~a_3} =0~, \qquad \epsilon^{b_1\ldots b_N} \epsilon_{a_1\ldots a_N} M^{a_1}_{~b_1} M^{a_2}_{~b_2}=0~,
\eea
where the indices $a_1, a_2, \ldots,a_N, b_1,\ldots, b_N = 1, \ldots, N$ are the fundamental indices of $SU(N)$.  Note that the first relations, which indicate that $M$ is a nilpotent matrix, transform in the representation $[1,0,\ldots,0,1]+[0,\ldots,0]$ of $SU(N)$.   The second relations transform in the representation $[0,1,0,\ldots,0,1,0]$ of $SU(N)$.

%

\subsection{Two instantons}
The generators of the reduced moduli space of two $G$ instantons on $\BC^2$ transform under the global symmetry $SU(2) \times G$ as stated in ~\tref{tab:genTwoInsts}.
\begin{table}[H]
\begin{center}
\begin{tabular}{|c|c|}
\hline
Order & Representation of $SU(2) \times G$ \\
\hline
2 & [2; {\bf 0}]+[0; {\bf Adj}] \\
3&  \hspace{1.1cm} [1; {\bf Adj}] \\
\hline
\end{tabular}
\end{center}
\caption{Generators of the reduced moduli space of two $G$ instantons on $\BC^2$ and how they transform under the global symmetry $SU(2) \times G$.}
\label{tab:genTwoInsts}
\end{table}%

There is one relation at order 4 in the representation $[0; {\bf 0}]$ of $SU(2) \times G$.  Explicitly, this relation can be written as 
\bea
\det X +c {\rm Tr} (M^2) =0~, 
\eea 
where $X$ and $M$ are the generators at order 2 in the representation $[2;0]$ and $[0;{\bf Adj}]$ of $SU(2)\times G$ respectively; the determinant corresponds to the $SU(2)$ group and ${\rm Tr}$ denotes the trace in the adjoint representation of $G$; the constant $c$ depends on the group $G$.

There are also relations at order $5$ in the representation $[1; {\bf Adj}] + [1; {\rm Sym}^2 {\bf Adj}-2 \cdot {\Adj}]$, where the notation ${\rm Sym}^2 {\bf Adj}-2 \cdot {\Adj}$ is as before.  This result agrees with the plethystic logarithm of the expression (3.11) in \cite{Keller:2012da}.

\subsection{Three instantons}
The generators of the reduced moduli space of three $G$ instantons on $\BC^2$ transform under the global symmetry $SU(2) \times G$ as stated in ~\tref{tab:gen3insts}.
\begin{table}[H]
\begin{center}
\begin{tabular}{|c|c|}
\hline
Order & Representation of $SU(2) \times G$ \\
\hline
2 & [2; {\bf 0}]+ [0; {\bf Adj}] \\
3&  [3; {\bf 0}]+ [1; {\bf Adj}]  \\
4 & \hspace{1.3cm} [2; {\bf Adj}]\\
\hline
\end{tabular}
\end{center}
\caption{Generators of the reduced moduli space of three $G$ instantons on $\BC^2$ and how they transform under the global symmetry $SU(2) \times G$.}
\label{tab:gen3insts}
\end{table}%

There is a set of relations at order 5 in the representation $[1; {\bf 0}]$ of $SU(2) \times G$.  Explicitly, this relation can be written as 
\bea
M_a G^\alpha_a  =0~, 
\eea 
where $M_a$ are the generators of the moduli space at order 2 in the representation $[0; {\bf Adj}]$ and $G^\alpha_a$ are the generators at order $3$ in the representation $[1; {\bf Adj}]$.  Here $a=1, \ldots, {\rm dim}\; G$ is an adjoint index of $G$ and $\alpha=1,2$ is an $SU(2)$ fundamental index.

\subsubsection*{Analytical properties of Hilbert series for three instantons} 
As discussed around (2.4) of \cite{Hanany:2012dm}, the Hilbert series of three $G$ instantons on $\BC^2$ shares certain analytical properties with the third symmetric power of the Hilbert series of one $G$ instanton on $\BC^2$, namely
\be \label{analprop}
\begin{split}
&\lim_{x \rightarrow a}(1-t^2 x^{-2})(1-t^3 x^{-3})   \widetilde{g}_{{\rm Sym}^3 \CM_{1,G}} (t; x; \vec u) \\
& = \lim_{x \rightarrow a} (1-t^2 x^{-2})(1-t^3 x^{-3})  \widetilde{g}_{3,G} (t; x; \vec u) ~,  \quad \text{with $a= \pm t, \; e^{\pm 2 \pi i/3} t$}~,
\end{split}
\ee
where a tilde denotes the Hilbert series of a {\it reduced} instanton moduli space, $x$ is the fugacity of $SU(2)$, and $\vec u$ denote the fugacities of the group $G$, and the third symmetric power is given by  
\be \label{sym3}
\begin{split}
\widetilde{g}_{\Sym^3 \CM_{1,G}} (t, x, \vec u) 
&= \frac{1}{6}  \Bigg[ \frac{1}{(1-t x^{\pm1})^2} \widetilde{g}_{1,G} (t, \vec u)^3 +3 \frac{1}{1-t^2 x^{\pm2}}  \widetilde{g}_{1,G} (t, \vec u) \widetilde{g}_{1,G} (t^2, \vec u^2)  \\ 
& \hspace{1cm} +2 \frac{1-t x^{\pm1}}{1-t^3 x^{\pm3}} \widetilde{g}_{1,G} (t^3, \vec u^3) \Bigg]~.
\end{split}
\ee
Explicitly, \eref{analprop} can be rewritten as follows:
\be \label{limsym3}
\begin{split}
 \lim_{x \rightarrow t} (1-t^2 x^{-2})(1-t^3 x^{-3}) \widetilde{g}_{3,G} (t; x; \vec u) &= \frac{\widetilde{g}_{1,G} (t, \vec u)^3}{(1-t^2)^{2}}~,    \\
 \lim_{x \rightarrow -t} (1-t^2 x^{-2})(1-t^3 x^{-3}) \widetilde{g}_{3,G} (t; x;  \vec u) &=\frac{\widetilde{g}_{1,G} (t, \vec u) \widetilde{g}_{1,G} (t^2, \vec u^2) }{1-t^4}~,   \\ 
 \lim_{x \rightarrow \omega t} (1-t^2 x^{-2})(1-t^3 x^{-3}) \widetilde{g}_{3,G} (t; x; \vec u) &=   \frac{1-\omega t^2}{1-t^6} \widetilde{g}_{1,G} (t^3, \vec u^3) ~, \quad \omega = e^{ \pm \frac{2 \pi i}{3}}~.
\end{split}
\ee
The properties \eref{limsym3} together with the fact that the numerator of the unrefined Hilbert series $\widetilde{g}_{3,G} (t; x=1; \vec u = {\bf 1})$ is palindromic can be used to check our results on the Hilbert series of three instantons.  

Let us demonstrate this for the case of $3$ $G_2$ instantons.  The numerator of the unrefined Hilbert series \eref{3G2HSur} is palindromic.  In order to make use of \eref{limsym3}, one needs to compute a refined Hilbert series at least with respect to $x$.   To keep the presentation brief, let us report the result for $3$ $G_2$ instantons up to order $t^9$:
\bea
& \tilde{g}_{3, G_2}(t; x; \vec u ={\bf 1}) \nn \\
&= 1+t^2 \left(x^2+\frac{1}{x^2}+15\right)+t^3 \left(x^3+\frac{1}{x^3}+15 x+\frac{15}{x}\right)+t^4 \left(x^4+\frac{1}{x^4}+29 x^2+\frac{29}{x^2}+135\right) \nn \\
& \quad +t^5 \left(x^5+\frac{1}{x^5}+30 x^3+\frac{30}{x^3}+240 x+\frac{240}{x}\right)+t^6 \Big(2 x^6+\frac{2}{x^6}+44 x^4+\frac{44}{x^4} \nn \\
& \quad +437 x^2+\frac{437}{x^2}+1102\Big)+t^7 \left(x^7+\frac{1}{x^7}+44 x^5+\frac{44}{x^5}+542 x^3+\frac{542}{x^3}+2292 x+\frac{2292}{x}\right)\nn \\
& \quad +t^8 \left(2 x^8+\frac{2}{x^8}+59 x^6+\frac{59}{x^6}+739 x^4+\frac{739}{x^4}+4232 x^2+\frac{4232}{x^2}+7964\right) \nn \\
& \quad +t^9 \left(2 x^9+\frac{2}{x^9}+59 x^7+\frac{59}{x^7}+844 x^5+\frac{844}{x^5}+5962 x^3+\frac{5962}{x^3}+17057 x+\frac{17057}{x}\right) \nn \\
& \quad + \ldots~,
\eea
and for $1$ $G_2$ instanton we have
\bea
\tilde{g}_{1, G_2}(t; \vec u ={\bf 1}) &= \sum_{p=0}^\infty \dim_{G_2} [p,0] t^{2p} \nn \\
&= 1 + 14 t^2 + 77 t^4 + 273 t^6 + 748 t^8 + 1729 t^{10}+\ldots~.
\eea
These can be substituted in \eref{limsym3} and the agreement on each equality can be obtained perturbatively up to order $t^4$.

\subsection{Higher instanton numbers}
Explicit computations reveal that the generators of the reduced moduli space of five $G$ instantons on $\BC^2$ transform under the global symmetry $SU(2) \times G$ as stated in \tref{tab:gen5insts}. 
\begin{table}[H]
\begin{center}
\begin{tabular}{|c|c|}
\hline
Order & Representation of $SU(2) \times G$ \\
\hline
2 & [2; {\bf 0}]+ [0; {\bf Adj}] \\
3&  [3; {\bf 0}]+[1; {\bf Adj}]  \\
4 &  [4; {\bf 0}]+[2; {\bf Adj}]\\
5 & [5; {\bf 0}]+[3; {\bf Adj}] \\
6 & \hspace{1.1cm} [4; {\bf Adj}] \\
\hline
\end{tabular}
\end{center}
\caption{Generators of the reduced moduli space of 5 $G$ instantons on $\BC^2$ and how they transform under the global symmetry $SU(2) \times G$.}
\label{tab:gen5insts}
\end{table}%

\subsection{Generators of the reduced instanton moduli spaces}

The data gathered in the previous subsection leads us to conjecture that the reduced moduli space of $k$ $G$ instantons on $\BC^2$ is generated by two sets of holomorphic functions transforming in:
\begin{enumerate}
\item representations $[p;\vec 0]$ of $SU(2) \times G$ at order $p$, for all $2 \leq p \leq k$;
\item representations $[p;{\bf Adj}]$ of $SU(2) \times G$ at order $p+2$, for all $0 \leq p \leq k-1$.
\end{enumerate}
These two sets of generators can be systematically understood from the Coulomb branch viewpoint, as we now explain. 


The generators transforming in the representation $[p;\vec 0]$ are all monopole operators. To describe them, it is useful to introduce a class of monopole operators that are obtained by embedding $U(k)$ monopoles into the $\prod_{i=0}^r U(k a_i^\vee) $ gauge group of the quiver. Let $M=\diag(m_1,m_2,\dots,m_k)$ be a $U(k)$ magnetic charge and consider the monopole operators of magnetic charge
\be\label{embedding_general}
m_{-1}=0~, \qquad 
m_i= M \otimes \frac{a_i}{a_i^\vee}~\unit_{a_i^\vee}
~, \quad i=0,\dots,r~,
\ee
generalizing \eqref{monopole_generating_C}.
The dimension of these monopole operators can be easily computed: the contributions of nodes and edges of the affine Dynkin diagram cancel out because the quiver is balanced, while the edge attached to the over-extended node yields $\Delta=\frac{1}{2}\sum_{i=1}^k |m_i|$.%
\footnote{For instance, for $F_4$ we compute \\~\\
$\Delta = \frac{1}{2}\sum\limits_i |m_i| -\frac{1}{2}\sum\limits_{i,j}|m_i-m_j|(2+6+12+4)- \sum\limits_{i<j}|m_i-m_j|(1+4+9+8+2)= \frac{1}{2}\sum\limits_i |m_i|~.$
}
Taking into account the charge under the topological symmetry group, the monopole operators \eqref{embedding_general} appear in the HS with weight $x^{\sum_i m_i} t^{\sum_i |m_i|}$.

Next, let
\be\label{sigma}
\sigma_{p,\ell}\equiv \diag(1^{p-\ell},(-1)^\ell)~,\qquad \ell=0,1,\dots, p
\ee 
be a $p\times p$ diagonal matrix with entries equal to $\pm 1$, which may be thought of as a collection of spins $\pm\frac{1}{2}$ for an abstract $SU(2)$.  This abstract $SU(2)$ is identified with the $SU(2)_x$ global symmetry of the instanton moduli space by specializing the matrix $M$ in \eqref{embedding_general} to  
\be\label{sigma_in_M}
M= \diag(\sigma_{p,\ell}, 0^{k-p}) 
\ee
up to Weyl reflections, where $p=1,2,\dots,k$ so that the $p\times p$ matrix $\sigma_{p,\ell}$ fits in the $k\times k$ matrix $M$. The case $p=1$ gives the generators of the center of the instanton, that was discussed in \eqref{monopole_generating_C}. The cases $p=2,\dots,k$ yield the generators of the reduced instanton moduli space in the representations $[p;\vec 0]$ of $SU(2) \times G$. Indeed the monopole operators of magnetic charge \eqref{embedding_general}, \eqref{sigma_in_M} appear in the HS with weights $x^{p-2\ell} t^p$. As $\ell=0,1,\dots,p$ at fixed $p$, they span the representation $[p;\vec 0]$ of $SU(2) \times G$.

One can similarly identify the generators at order $p+2$ transforming in the representation $[p;{\bf Adj}]$, where $p=0,1,\dots,k-1$. Let us first restrict to the positive roots $\vec \alpha$ of $G$, keeping all weights of $SU(2)$ representations. The generators are monopole operators of magnetic charges
\be\label{mixed_generators}
m_{-1}=0~, \qquad 
m_i = \diag\left(R_i^{(\vec \alpha)}, \sigma_{p,\ell}\otimes \frac{a_i}{a_i^\vee}\unit_{a_i^\vee}, 0^{\frac{a_i}{a_i^\vee}(k-1-p)}\right)
~, \quad i=0,\dots,r~,
\ee
where $R_i^{(\vec \alpha)}$ is an $a_i^\vee\times a_i^\vee$ diagonal matrix whose elements are tabulated in Appendix \ref{sec:enhance} for non-simply laced groups and can be found in \cite{Bashkirov:2010hj} for simply-laced groups. $R_0^{(\vec \alpha)}$ is always zero. Note that $p$ necessarily runs from $0$ to $k-1$. The contribution of $R_i^{(\vec \alpha)}$ to the topological charge of the monopole operator reproduces the positive root $\vec \alpha$ of $G$, whereas  $\sigma_{p,\ell}$ is responsible for the $SU(2)$ weight ${p-2\ell}$ as above. 
For negative roots of $G$, $R_i^{(\vec \alpha)}$ is replaced by its negative.  For the Cartan elements of $G$, $R_i^{(\vec \alpha)}$ are set to zero and the monopole operators are dressed by the classical field at the $i$-th node of the Dynkin diagram of $G$.

\subsection{Monopole operators and global symmetries}\label{sec:mon_global_symm}

The global symmetry group acting on the Coulomb branch of a $3d$ $\cN=4$ superconformal field theory takes the form $SU(2)_C \times G_J$. $SU(2)_C$ is an $R$-symmetry which rotates the triplet of complex structures of the hyperK\"ahler manifold. The holomorphic functions with respect to a fixed complex structure that are counted by the HS are highest weights of $SU(2)_C$ representations. The associated fugacity is $t$. On the other hand, $G_J$ commutes with the supercharges. A subgroup of $G_J$ is manifest in the UV Lagrangian of the gauge theory: it consists of the topological symmetry group which is generated by the topologically conserved currents $J_i=\ast \Tr F_i$, where $F_i$ are the field strength 2-forms of the $i$-th $U(N_i)$ gauge group. More generally, the topological symmetry is the center $\cZ(\cG^\vee)$ of the dual of the gauge group. The topological symmetry group, which is $U(1)^{r+1}$ for the theories considered in this paper, acts on monopole operators. The associated fugacities are $z_i$.

At the IR fixed point of a three-dimensional gauge theory, the manifest topological symmetry group can enhance to a non-abelian symmetry group $G_J$. The conserved currents of the hidden symmetry are monopole operators. In a $3d$ $\cN=4$ superconformal field theory, conserved currents sit in the same multiplet as dimension $\Delta=1$ chiral operators \cite{Gaiotto:2008ak} (see also \cite{Bashkirov:2010kz,Bashkirov:2010hj}). Thus the non-R global symmetry can be deduced from the Hilbert series: the order $t^2$ term gives the adjoint representation of $G_J$. 

Applying this strategy to the quivers whose Coulomb branches are the moduli spaces of instantons, one can see that the global non-R symmetry enhances from $U(1)^{r+1}$ to $G_J=SU(2)\times G$ for $k=1$ instanton and to $G_J=SU(2)\times SU(2)\times G$ for $k>1$ instantons, as we now explain. 

The maximal torus $U(1)^r$ of $G$ is the manifest topological symmetry associated to the nodes of the Dynkin diagram of $G$ in the quiver. The $\Delta=1$ states counted by the Hilbert series are $\Tr \Phi_i$, $i=1,\dots,r$, where $\Phi_i$ is the adjoint chiral multiplet in the $\cN=4$ vector multiplet of the $i$-th gauge group. The global symmetry enhancement is due to dimension $1$ monopole operators in one-to-one correspondence with the roots of $G$. For positive roots $\vec \alpha$, these dimension $1$ monopole operators take the form
\be\label{positive_roots_G}
m_{-1}=0~, 
\quad m_i = \diag\left(R_i^{(\vec \alpha)}, 0^{\frac{a_i}{a_i^\vee}(k-1)}\right)
~, \quad i=0,\dots,r~,
\ee
where $R_i^{(\vec \alpha)}$ is an $a_i^\vee\times a_i^\vee$ diagonal matrix whose elements are tabulated in Appendix \ref{sec:enhance} for non-simply laced groups and can be found in \cite{Bashkirov:2010hj} for simply-laced groups. Note that $R_0^{(\vec \alpha)}$ is always zero. The topological charge $\Tr R_i^{(\vec \alpha)}$ of the monopole operator is the component of the positive root $\vec \alpha$ of $G$ along the $i$-th simple root of $G$. For instance, for $G=SU(N+1)$, the positive roots are ${\vec \alpha}_{ij}=\sum_{p=i}^{j-1} {\vec \gamma}_p$, with ${\vec \gamma}_p$ the simple roots and $1\leq i<j\leq N$. Then $R_p^{(\vec \alpha_{ij})}=(1)$ if $i\leq p <j$ and $R_p^{(\vec \alpha_{ij})}=(0)$ otherwise. The negative roots of $G$ are obtained by flipping sign to the magnetic charges \eqref{positive_roots_G}.

Next we explain the $SU(2)$ groups. The $SU(2)$ symmetry that is present for any instanton number $k$ acts on the two complex variables parametrizing the center of the instanton configuration, namely the monopole operators of magnetic charges $\pm 1$ times \eqref{monopole_generating_C}. The squares of those monopole generators, corresponding to magnetic charges $\pm 2$ times \eqref{monopole_generating_C}, provide the roots of $SU(2)$; the classical field $\sum_{i=0}^r \frac{a_i}{a_i^\vee} \Tr \Phi_i$ associated to the remaining $U(1)$ topological symmetry provides the Cartan element of $SU(2)$.

For instanton number $k>1$ there is an additional $SU(2)$ which acts on the reduced moduli space of instantons. The adjoint representation of this additional $SU(2)$ is spanned by monopole operators of the form \eqref{embedding_general}, \eqref{sigma_in_M}, where $p=2$ in \eqref{sigma}.

Note that the characters of the adjoint representations of the two $SU(2)$ factors that appear in the HS at order $t^2$ involve the same fugacity $x$ for the diagonal $SU(2)$ defined in \eref{x_definition}. Since the symmetry is $SU(2)\times SU(2)$, it should be possible to further refine the Hilbert series of the instanton moduli space and distinguish the two $SU(2)$ factors. However, for one of the $SU(2)$ groups, not even the Cartan subalgebra is manifest, but rather it is generated by a monopole operator.  This difficulty can be circumvented because the center of the instanton is factored in the Hilbert series and is represented by a free twisted hypermultiplet.  One can always {\it a posteriori} assign different fugacities to the two $SU(2)$ factors (cf. (3.3) of \cite{Gaiotto:2012uq}), modifying \eref{nonfactorisableModSpace} as follows:
\bea
g_{k,G}(t,x_1, x_2 ,\vec{u})=\frac{1}{(1-tx_1)(1-tx_1^{-1})} {\tilde g}_{k,G}(t,x_2,\vec{u})~.
\eea



\section{Conclusions}\label{sec:concl}

In this paper we have proposed a simple formula for the Hilbert series of moduli spaces of pure Yang-Mills instantons, which arise as Coulomb branches of three-dimensional $\cN=4$ generalized quiver gauge theories whose quiver diagrams are given by over-extended Dynkin diagrams. A natural modification of the monopole formula for the Coulomb branch Hilbert series introduced in \cite{Cremonesi:2013lqa} allows us to uniformly study instantons in all simple Lie groups, including the non-simply laced ones. We have successfully tested our proposal against previous works for one and two instantons and obtained new results for higher instanton numbers. General features of the moduli spaces of instantons can be systematically deduced from our formalism. 
It would be interesting to derive the explicit ring structure of the moduli spaces by a careful analysis of monopole operators.

Our work leaves some natural open questions. Firstly, it would be nice to derive our formula from a path integral by folding the appropriate simply laced quiver via an outer automorphism group. This would help to understand the Higgs branch of such quivers and compute superconformal indices \cite{Kim:2009wb,Imamura:2011su,Kapustin:2011jm}.  
Secondly, the Coulomb branch formalism should also allow for the computation of the hyperK\"ahler metric on the moduli spaces of instantons \cite{Intriligator:1996ex,deBoer:1996mp}. Indeed, formulae (4.2)--(4.4) in \cite{deBoer:1996mp} could be generalized to non-simply laced quivers by inserting the multiplicity $\lambda$ in the matter contribution to the metric in analogy with \eqref{dimension_hyper_guess}. For classical groups, this suggestion can be tested against the metric obtained from the hyperK\"ahler quotient in the Higgs branch of the corresponding ADHM quiver.
Finally, it would be interesting to generalize the Coulomb branch construction of this paper to instantons on ALE spaces \cite{deBoer:1996mp, Porrati:1996xi, Dey:2014tka}.

\acknowledgments
We thank Jan Troost for useful discussions and particularly Alberto Zaffaroni for a close collaboration and invaluable insights over the years. The following institutes and workshops are gratefully acknowledged for hospitality and partial support: KITP Program on New Methods in Nonperturbative Quantum Field Theory, supported partly by the National Science Foundation under Grant No. PHY11-25915 (AH and NM); University of Texas at Austin (NM); Perimeter Institute (NM); Gauge theories: quivers, tilings and Calabi-Yaus, ICMS Edinburgh (SC, AH and NM); Localisation and the gauge/gravity duality, King's College London (SC, AH and NM); Imperial College London and Queen Mary University of London (NM); Carg\`ese Summer Institute 2014 and the traveling grant from the Weizmann Institute of Science via Zohar Komargodski (NM); Exact Quantum Fields and the Structure of M-theory, University of Crete (NM); String Theory and Holography Summer School at IST Lisbon and Porto (SC and GF);
the Simons Center for Geometry and Physics and the 2014 Workshop (AH and NM); the CERN-Korea Theory Collaboration funded by National Research Foundation (Korea) and the workshop Exact Results in SUSY Gauge Theories in Various Dimensions at CERN (SC and NM); the XLIV\`eme Institut d'\'Et\'e at ENS Paris (SC).

\appendix
\section{The Hilbert series for $k$ $USp(2N)$ instantons for odd $k$ via Higgs branch} \label{sec:USp2NHiggs}
For higher number of instantons, the Hilbert series can be computed more easily from the Higgs branch of the ADHM quiver.  In particular for $k$ odd the Hilbert series is given by
\be
\begin{split}
&g_{k, USp(2N)}(t; x; \vec u)  \\
&= \frac{1}{2} \sum_{\omega  = \pm 1} \int {\rm d} \mu_{SO(k)} (\vec z)  \PE \Big[  \omega \chi^{USp(2N)}_{[1,0,\ldots,0]} (\vec u) \chi^{SO(k)}_{[1,0,\ldots,0]} (\vec z)t  \\
&\quad + (x+x^{-1}) (\chi^{SO(k)}_{[2,0,\ldots,0]}(\vec z)+1)t - t^2  \chi^{SO(k)}_{[0,1,0,\ldots,0]} (\vec z)\Big]~, \qquad \text{($k$ odd)}
\end{split}
\ee
where for $SO(k)$, the Dynkin labels $[1,0,\ldots,0]$, $[2,0,\ldots,0]$, $[0,1,0,\ldots,0]$ denotes the vector, the symmetric traceless, and the adjoint representations respectively. Here $\omega$ corresponds to the parity action $\pm 1$ of $O(k) = SO(k) \times \{ \pm 1\}$ for odd $k$. 
The Haar measure of $SO(2k+1)$ is given by
\be
\begin{split}
& \int {\rm d} \mu_{SO(2k+1)} (\vec z)  \\
 & = \oint_{|z_1|=1} \frac{{\rm d} z_1}{ 2 \pi i z_1} \cdots \oint_{|z_k|=1} \frac{{\rm d} z_k}{2 \pi i z_k} \prod_{1\leq i < j \leq k} \left( 1-z_i z_j\right)\left( 1-z_i z_j^{-1}\right) \prod_{m=1}^k (1-z_m)~,
\end{split}
\ee
where the adjoint representation is taken as
\bea
\chi^{SO(2k+1)}_{[0,1,0,\ldots,0]} (\vec z) = \sum_{1\leq i < j \leq k} \left(z_i z_j + z_i z_j^{-1}\right) + \sum_{m=1}^k z_m
\eea
The Hilbert series for the reduced moduli space of instantons is then given by
\bea
{\tilde g}_{k, USp(2N)}(t; x; \vec u) = (1-t x)(1-t x^{-1}) g_{k, USp(2N)}(t; x; \vec u)~. \
\eea
\section{Monopole operators and symmetry enhancement} \label{sec:enhance}
\subsection{$G_2$}
The relevant diagram for $k$ $G_2$ instantons is depicted below.
\bea
{\blue \node{}{1}}-\Node{{\vec \alpha}_0}{k}-\node{{\vec \alpha}_1}{2k}\Rrightarrow \node{{\vec \alpha}_2}{k} \ \tikz[na]\node(quivG2){};
\eea
where the simple roots ${\vec \alpha}_0, {\vec \alpha}_1, {\vec \alpha}_2$ are indicated above the nodes.  The positive roots of $G_2$ are of the form $c_1 {\vec \alpha}_1 + c_2 {\vec \alpha}_2$, with $(c_1, c_2)$ listed in \tref {tab:posrootsG2mono}.  For each positive root, we tabulate the monopole operators associated with it.

\begin{table}[H]
\begin{center}
\begin{tabular}{|c||c|c|c|}
\hline
Positive root  & $R^{(\vec \alpha_0)}$ & $R^{(\vec \alpha_1)}$   & $R^{(\vec \alpha_2)}$ \\
\hline
$(1,0)$& $(0)$  & $(1,0)$ & $(0)$ \\
\hline
$(0,1)$& $(0)$  & $(0,0)$ & $(1)$\\
\hline
$(1,1)$& $(0)$  & $(1,0)$ & $(1)$\\
\hline
$(1,2)$& $(0)$  & $(1,0)$ & $(2)$\\
\hline
$(1,3)$& $(0)$  & $(1,0)$ & $(3)$\\
\hline
$(2,3)$& $(0)$  & $(1,1)$ &   $(3)$  \\
\hline
\end{tabular}
\end{center}
\caption{Magnetic charges $R^{(\vec \alpha)}_i$ of the monopole operators that contribute to each positive root $\vec \alpha$ of $G_2$ for $k=1$ instanton.}
\label{tab:posrootsG2mono}
\end{table}%

\subsection{$F_4$}
The relevant diagram for $k$ $F_4$ instantons is depicted below.
\bea
{\blue \node{}{1}}- \Node{\vec \alpha_0}{k}-\node{\vec \alpha_1}{2k}-\node{\vec \alpha_2}{3k}\Rightarrow\node{\vec \alpha_3}{2k}-\node{\vec \alpha_4}{k} \quad\tikz[na]\node(quivF4){};
\eea
The $24$ positive roots of $F_4$ are of the form $\sum_{i=1}^4 c_i {\vec \alpha}_i$, with $(c_1, \ldots, c_4)$ listed in \tref {tab:posrootsF4mono}.  For each positive root, we tabulate the monopole operators associated with it.
\begin{table}[htdp]
\begin{center}
{\small
\begin{tabular}{|c||c|c|c|c|c|}
\hline
Positive root    & $R^{(\vec \alpha_0)}$ & $R^{(\vec \alpha_1)}$   & $R^{(\vec \alpha_2)}$ & $R^{(\vec \alpha_3)}$ & $R^{(\vec \alpha_4)}$   \\
\hline
$(1,0,0,0)$ & $(0)$ & $(1,0)$ & $(0,0,0)$ & $(0,0)$ & $(0)$\\
\hline
$(0,1,0,0)$ & $(0)$ & $(0,0)$ & $(1,0,0)$ & $(0,0)$ & $(0)$\\
\hline
$(0,0,1,0)$ & $(0)$ & $(0,0)$ & $(0,0,0)$ & $(1,0)$ & $(0)$\\
\hline
$(0,0,0,1)$ & $(0)$ & $(0,0)$ & $(0,0,0)$ & $( 0,0)$ & $(1)$\\
\hline
$(1,1,0,0)$ & $(0)$ & $(1,0)$ & $(1,0,0)$ & $( 0,0)$ &$(0)$ \\
\hline
$(0,1,1,0)$& $(0)$  & $(0,0)$ & $(1,0,0)$ & $(1,0)$ &$(0)$ \\
\hline
$(0,0,1,1)$& $(0)$  & $(0,0)$ & $(0,0,0)$ & $(1,0)$ &$(1)$ \\
\hline
$(0,1,1,1)$& $(0)$  & $(0,0)$ & $(1,0,0)$ & $(1,0)$ &$(1)$ \\
\hline
$(1,1,1,0)$& $(0)$  & $(1,0)$ & $(1,0,0)$ & $(1,0)$ &$(0)$ \\
\hline
$(1,1,1,1)$& $(0)$  & $(1,0)$ & $(1,0,0)$ & $(1,0)$ &$(1)$ \\
\hline
$(0,1,2,0)$& $(0)$  & $(0,0)$ & $(1,0,0)$& $(2,0)$& $(0)$\\
\hline
$(1,1,2,0)$& $(0)$  & $(1,0)$ &   $(1,0,0)$ &   $(2,0)$ &   $(0)$ \\
\hline
$(0,1,2,1)$& $(0)$  & $(0,0)$ &   $(1,0,0)$ &   $(2,0)$ &   $(1)$\\
\hline
$(1,2,2,0)$& $(0)$  & $(1,0)$ &   $(1,1,0)$ &   $(2,0)$ &   $(0)$\\
\hline
$(1,1,2,1)$& $(0)$  & $(1,0)$ &   $(1,0,0)$ &   $(2,0)$ &$(1)$ \\
\hline
$(0,1,2,2)$& $(0)$  & $(0,0)$ &   $(1,0,0)$ & $(2,0)$& $(2)$ \\
\hline
$(1,2,2,1)$& $(0)$  & $(1,0)$ &   $(1,1,0)$ &$(2,0)$& $(1)$  \\
\hline
$(1,1,2,2)$& $(0)$  & $(1,0)$ &   $(1,0,0)$ & $(2,0)$& $(2)$ \\
\hline
$(1,2,3,1)$& $(0)$  & $(1,0)$ &   $(1,1,0)$& $(2,1)$& $(1)$  \\
\hline
$(1,2,2,2)$& $(0)$  & $(1,0)$ &   $(1,1,0)$& $(2,0)$& $(2)$  \\
\hline
$(1,2,3,2)$& $(0)$  & $(1,0)$ &   $(1,1,0)$ & $(2,1)$& $(2)$ \\
\hline
$(1,2,4,2)$& $(0)$  & $(1,0)$ &   $(1,1,0)$ & $(2,2)$& $(2)$ \\
\hline
$(1,3,4,2)$& $(0)$  & $(1,0)$ &   $(1,1,1)$  & $(2,2)$& $(2)$ \\
\hline
$(2,3,4,2)$& $(0)$  & $(1,1)$ &   $(1,1,1)$  & $(2,2)$& $(2)$ \\
\hline
\end{tabular}}
\end{center}
\caption{Magnetic charges $R^{(\vec \alpha)}_i$ of the monopole operators that contribute to each positive root $\vec \alpha$ of $F_4$ for $k=1$ instanton.}
\label{tab:posrootsF4mono}
\end{table}%

\subsection{$C_N$}
The relevant diagram for $k$ $USp(2N)$ instantons is depicted below.
\bea \label{quivvC}
{\blue \node{}{1}}-\Node{\vec \alpha_0}{k}\Rightarrow \underbrace{\node{\vec \alpha_1}{k}-\cdots-\node{\vec \alpha_{N-1}}{k}}_{N-1~\text{nodes}}\Leftarrow\node{\vec \alpha_{N}}{k}  \ \tikz[na]\node(quivC){};
\eea
where the simple roots are indicated above each node.
The positive roots of $USp(2N)$ are
\bea
\Delta_+ = \{ \vec e_i +\vec e_j \}_{i<j} \cup  \{ \vec e_i -\vec e_j \}_{i<j} \cup  \{ 2\vec e_i \}_{i=1}^N~,
\eea
where $\{ \vec e_i\}$ is the standard basis.
The simple roots of $USp(2N)$ can be written as
\be
\begin{split}
\vec \alpha_\ell &= \vec e_\ell - \vec e_{\ell+1}~, \qquad 1 \leq \ell \leq N-1~,  \\
\vec \alpha_N &= 2 \vec e_N~.
\end{split}
\ee
The positive roots can be written in terms of the simple roots as
\be
\begin{split}
\vec e_i - \vec e_j &= \sum_{\ell =i}^{j-1} \alpha_\ell~, \\
 2 \vec e_i &= 2\sum_{\ell =i}^{N-1} \vec \alpha_\ell + \vec \alpha_N~,  \\
  \vec e_i +\vec e_j &= \sum_{\ell =i}^{j-1} \vec \alpha_\ell + 2\sum_{\ell =j}^{N-1} \vec \alpha_\ell + \vec \alpha_N~. 
\end{split}
\ee

The magnetic charges $R^{(\vec \alpha)}_i$ of the monopole operators that contribute to each positive root $\vec \alpha$ of $C_N$ for $k=1$ instanton are as follows:
\bi
\item  $\vec e_i - \vec e_j$: $(1)$ from nodes $\vec \alpha_p$ with $1 \leq p \leq j-1$, and $(0)$ from other nodes.
\item  $2 \vec e_i$: $(0)$ from node $\vec \alpha_0$, $(2)$ from nodes $\vec \alpha_p$ with $ i \leq p \leq N-1$, and $(1)$ from node $\vec \alpha_N$.
\item $\vec e_i +\vec e_j$: $(0)$ from node $\vec \alpha_0$, $(1)$ from node $\vec \alpha_p$ with $1 \leq p \leq j-1$, $(2)$ from node $\vec \alpha_q$ with $j \leq q \leq N-1$, and $(1)$ from node $\vec \alpha_N$.
\ei

\subsection{$B_N$}
The relevant diagram for $k$ $SO(2N+1)$ instantons is depicted below.
\bea
{\blue \node{}{1}}- \Node{\vec \alpha_0}{k}-\node{\ver{\vec \alpha_1}{k}}{\substack{2k \\ {\vec \alpha_2}}}-\underbrace{\node{\vec \alpha_3}{2k}-\cdots-\node{\vec \alpha_{N-1}}{2k}}_{N-3~\text{nodes}}\Rightarrow\node{\vec \alpha_N}{k}  \ \tikz[na]\node(B1){};
\eea
where the simple roots are indicated at each node. 
The positive roots of $SO(2N+1)$ are
\bea
\Delta_+ = \{ \vec e_i +\vec e_j \}_{i<j} \cup  \{ \vec e_i -\vec e_j \}_{i<j} \cup  \{ \vec e_i \}_{i=1}^N~,
\eea
where $\{ \vec e_i\}$ is the standard basis.
The simple roots of $SO(2N+1)$ can be written as
\be
\begin{split}
\vec \alpha_\ell &= \vec e_\ell - \vec e_{\ell+1}~, \qquad 1 \leq \ell \leq N-1~, \\
\vec \alpha_N &=  \vec e_N~.
\end{split}
\ee
The positive roots can be written in terms of the simple roots as
\be
\begin{split}
\vec e_i - \vec e_j &= \sum_{\ell =i}^{j-1} \alpha_\ell~, \\
  \vec e_i &= \sum_{\ell =i}^{N} \vec \alpha_\ell ~, \\
  \vec e_i +\vec e_j &= \sum_{\ell =i}^{j-1} \vec \alpha_\ell + 2\sum_{\ell =j}^{N} \vec \alpha_\ell ~. 
\end{split}
\ee

The magnetic charges of the monopole operators that contribute to each positive root $\vec \alpha$ of $B_N$ for any instanton number are as follows:
\bi
\item $\vec e_i$: $(1, \vec 0)$ from the nodes $\vec \alpha_p$ with $i \leq p \leq N$, and $(\vec 0)$ from other nodes.
\item  $\vec e_i - \vec e_j$: $(1, \vec 0)$ from the nodes $\vec \alpha_p$ with $i \leq p \leq j-1$, and $(\vec 0)$ from other nodes.
\item $\vec e_i +\vec e_j$: $(1,\vec 0)$ from the nodes $\vec \alpha_p$ with $i \leq p \leq j-1$, $(1^2, \vec 0)$ from the nodes $\vec \alpha_q$ with $j \leq  q \leq N-1$, $(2)$ from the node $\vec \alpha_N$.
\ei

\bibliographystyle{ytphys}
\bibliography{ref_revised}

\providecommand{\href}[2]{#2}\begingroup\raggedright\begin{thebibliography}{10}

\bibitem{Belavin:1975fg}
A.~Belavin, A.~M. Polyakov, A.~Schwartz, and Y.~Tyupkin, ``{Pseudoparticle
  Solutions of the Yang-Mills Equations},''
\href{http://dx.doi.org/10.1016/0370-2693(75)90163-X}{{\em Phys.Lett.}
  {\bfseries B59} (1975) 85--87}.

\bibitem{PhysRevD.14.3432}
G.~'t~Hooft, ``Computation of the quantum effects due to a four-dimensional
  pseudoparticle,''
  \href{http://link.aps.org/doi/10.1103/PhysRevD.14.3432}{{\em Phys. Rev. D}
  {\bfseries 14} (Dec, 1976) 3432--3450}.

\bibitem{Atiyah:1978ri}
M.~Atiyah, N.~J. Hitchin, V.~Drinfeld, and Y.~Manin, ``{Construction of
  Instantons},''
\href{http://dx.doi.org/10.1016/0375-9601(78)90141-X}{{\em Phys.Lett.}
  {\bfseries A65} (1978) 185--187}.

\bibitem{Witten:1995gx}
E.~Witten, ``{Small instantons in string theory},''
  \href{http://dx.doi.org/10.1016/0550-3213(95)00625-7}{{\em Nucl.Phys.}
  {\bfseries B460} (1996) 541--559},
\href{http://arxiv.org/abs/hep-th/9511030}{{\ttfamily arXiv:hep-th/9511030
  [hep-th]}}.

\bibitem{Douglas:1995bn}
M.~R. Douglas, ``{Branes within branes},''
\href{http://arxiv.org/abs/hep-th/9512077}{{\ttfamily arXiv:hep-th/9512077
  [hep-th]}}.

\bibitem{Argyres:1996eh}
P.~C. Argyres, M.~R. Plesser, and N.~Seiberg, ``{The Moduli space of vacua of
  N=2 SUSY QCD and duality in N=1 SUSY QCD},''
  \href{http://dx.doi.org/10.1016/0550-3213(96)00210-6}{{\em Nucl.Phys.}
  {\bfseries B471} (1996) 159--194},
\href{http://arxiv.org/abs/hep-th/9603042}{{\ttfamily arXiv:hep-th/9603042
  [hep-th]}}.

\bibitem{Intriligator:1996ex}
K.~A. Intriligator and N.~Seiberg, ``{Mirror symmetry in three-dimensional
  gauge theories},'' \href{http://dx.doi.org/10.1016/0370-2693(96)01088-X}{{\em
  Phys.Lett.} {\bfseries B387} (1996) 513--519},
\href{http://arxiv.org/abs/hep-th/9607207}{{\ttfamily arXiv:hep-th/9607207
  [hep-th]}}.

\bibitem{Romelsberger:2005eg}
C.~Romelsberger, ``{Counting chiral primaries in N = 1, d=4 superconformal
  field theories},''
  \href{http://dx.doi.org/10.1016/j.nuclphysb.2006.03.037}{{\em Nucl.Phys.}
  {\bfseries B747} (2006) 329--353},
\href{http://arxiv.org/abs/hep-th/0510060}{{\ttfamily arXiv:hep-th/0510060
  [hep-th]}}.

\bibitem{Kinney:2005ej}
J.~Kinney, J.~M. Maldacena, S.~Minwalla, and S.~Raju, ``{An Index for 4
  dimensional super conformal theories},''
  \href{http://dx.doi.org/10.1007/s00220-007-0258-7}{{\em Commun.Math.Phys.}
  {\bfseries 275} (2007) 209--254},
\href{http://arxiv.org/abs/hep-th/0510251}{{\ttfamily arXiv:hep-th/0510251
  [hep-th]}}.

\bibitem{Gadde:2011uv}
A.~Gadde, L.~Rastelli, S.~S. Razamat, and W.~Yan, ``{Gauge Theories and
  Macdonald Polynomials},''
  \href{http://dx.doi.org/10.1007/s00220-012-1607-8}{{\em Commun.Math.Phys.}
  {\bfseries 319} (2013) 147--193},
\href{http://arxiv.org/abs/1110.3740}{{\ttfamily arXiv:1110.3740 [hep-th]}}.

\bibitem{Gaiotto:2009we}
D.~Gaiotto, ``{N=2 dualities},''
  \href{http://dx.doi.org/10.1007/JHEP08(2012)034}{{\em JHEP} {\bfseries 1208}
  (2012) 034},
\href{http://arxiv.org/abs/0904.2715}{{\ttfamily arXiv:0904.2715 [hep-th]}}.

\bibitem{Gaiotto:2012uq}
D.~Gaiotto and S.~S. Razamat, ``{Exceptional Indices},''
  \href{http://dx.doi.org/10.1007/JHEP05(2012)145}{{\em JHEP} {\bfseries 1205}
  (2012) 145},
\href{http://arxiv.org/abs/1203.5517}{{\ttfamily arXiv:1203.5517 [hep-th]}}.

\bibitem{Nakajima:2003pg}
H.~Nakajima and K.~Yoshioka, ``{Instanton counting on blowup. 1.},''
  \href{http://dx.doi.org/10.1007/s00222-005-0444-1}{{\em Invent.Math.}
  {\bfseries 162} (2005) 313--355},
\href{http://arxiv.org/abs/math/0306198}{{\ttfamily arXiv:math/0306198
  [math-ag]}}.

\bibitem{Nakajima:2005fg}
H.~Nakajima and K.~Yoshioka, ``{Instanton counting on blowup. II. K-theoretic
  partition function},''
\href{http://arxiv.org/abs/math/0505553}{{\ttfamily arXiv:math/0505553
  [math-ag]}}.

\bibitem{Keller:2012da}
C.~A. Keller and J.~Song, ``{Counting Exceptional Instantons},''
  \href{http://dx.doi.org/10.1007/JHEP07(2012)085}{{\em JHEP} {\bfseries 1207}
  (2012) 085},
\href{http://arxiv.org/abs/1205.4722}{{\ttfamily arXiv:1205.4722 [hep-th]}}.

\bibitem{Benvenuti:2010pq}
S.~Benvenuti, A.~Hanany, and N.~Mekareeya, ``{The Hilbert Series of the One
  Instanton Moduli Space},''
  \href{http://dx.doi.org/10.1007/JHEP06(2010)100}{{\em JHEP} {\bfseries 06}
  (2010) 100},
\href{http://arxiv.org/abs/1005.3026}{{\ttfamily arXiv:1005.3026 [hep-th]}}.

\bibitem{Keller:2011ek}
C.~A. Keller, N.~Mekareeya, J.~Song, and Y.~Tachikawa, ``{The ABCDEFG of
  Instantons and W-algebras},''
  \href{http://dx.doi.org/10.1007/JHEP03(2012)045}{{\em JHEP} {\bfseries 1203}
  (2012) 045},
\href{http://arxiv.org/abs/1111.5624}{{\ttfamily arXiv:1111.5624 [hep-th]}}.

\bibitem{Hanany:2012dm}
A.~Hanany, N.~Mekareeya, and S.~S. Razamat, ``{Hilbert Series for Moduli Spaces
  of Two Instantons},'' \href{http://dx.doi.org/10.1007/JHEP01(2013)070}{{\em
  JHEP} {\bfseries 1301} (2013) 070},
\href{http://arxiv.org/abs/1205.4741}{{\ttfamily arXiv:1205.4741 [hep-th]}}.

\bibitem{Feng:2007ur}
B.~Feng, A.~Hanany, and Y.-H. He, ``{Counting Gauge Invariants: the Plethystic
  Program},'' \href{http://dx.doi.org/10.1088/1126-6708/2007/03/090}{{\em JHEP}
  {\bfseries 03} (2007) 090},
\href{http://arxiv.org/abs/hep-th/0701063}{{\ttfamily arXiv:hep-th/0701063}}.

\bibitem{Nekrasov:2002qd}
N.~A. Nekrasov, ``{Seiberg-Witten prepotential from instanton counting},'' {\em
  Adv.Theor.Math.Phys.} {\bfseries 7} (2004) 831--864,
\href{http://arxiv.org/abs/hep-th/0206161}{{\ttfamily arXiv:hep-th/0206161
  [hep-th]}}.

\bibitem{Nekrasov:2004vw}
N.~Nekrasov and S.~Shadchin, ``{ABCD of instantons},''
  \href{http://dx.doi.org/10.1007/s00220-004-1189-1}{{\em Commun.Math.Phys.}
  {\bfseries 252} (2004) 359--391},
\href{http://arxiv.org/abs/hep-th/0404225}{{\ttfamily arXiv:hep-th/0404225
  [hep-th]}}.

\bibitem{Cremonesi:2013lqa}
S.~Cremonesi, A.~Hanany, and A.~Zaffaroni, ``{Monopole operators and Hilbert
  series of Coulomb branches of 3d N = 4 gauge theories},''
\href{http://arxiv.org/abs/1309.2657}{{\ttfamily arXiv:1309.2657 [hep-th]}}.

\bibitem{Gaiotto:2008ak}
D.~Gaiotto and E.~Witten, ``{S-Duality of Boundary Conditions In N=4 Super
  Yang-Mills Theory},'' {\em Adv. Theor. Math. Phys.} {\bfseries 13} (2009)
  721,
\href{http://arxiv.org/abs/0807.3720}{{\ttfamily arXiv:0807.3720 [hep-th]}}.

\bibitem{Razamat:2014pta}
S.~S. Razamat and B.~Willett, ``{Down the rabbit hole with theories of class
  S},''
\href{http://arxiv.org/abs/1403.6107}{{\ttfamily arXiv:1403.6107 [hep-th]}}.

\bibitem{Cremonesi:2014kwa}
S.~Cremonesi, A.~Hanany, N.~Mekareeya, and A.~Zaffaroni, ``{Coulomb branch
  Hilbert series and Hall-Littlewood polynomials},''
\href{http://arxiv.org/abs/1403.0585}{{\ttfamily arXiv:1403.0585 [hep-th]}}.

\bibitem{Cremonesi:2014vla}
S.~Cremonesi, A.~Hanany, N.~Mekareeya, and A.~Zaffaroni, ``{Coulomb branch
  Hilbert series and Three Dimensional Sicilian Theories},''
\href{http://arxiv.org/abs/1403.2384}{{\ttfamily arXiv:1403.2384 [hep-th]}}.

\bibitem{Borokhov:2002ib}
V.~Borokhov, A.~Kapustin, and X.-k. Wu, ``{Topological disorder operators in
  three-dimensional conformal field theory},'' {\em JHEP} {\bfseries 0211}
  (2002) 049,
\href{http://arxiv.org/abs/hep-th/0206054}{{\ttfamily arXiv:hep-th/0206054
  [hep-th]}}.

\bibitem{Kronheimer:1989zs}
P.~Kronheimer, ``{The Construction of ALE spaces as hyperKahler quotients},''
{\em J.Diff.Geom.} {\bfseries 29} (1989) 665--683.

\bibitem{KronheimerNakajima}
P.~Kronheimer and H.~Nakajima, ``Yang-mills instantons on ale gravitational
  instantons,'' \href{http://dx.doi.org/10.1007/BF01444534}{{\em Mathematische
  Annalen} {\bfseries 288} no.~1, (1990) 263--307}.

\bibitem{deBoer:1996mp}
J.~de~Boer, K.~Hori, H.~Ooguri, and Y.~Oz, ``{Mirror symmetry in
  three-dimensional gauge theories, quivers and D-branes},''
  \href{http://dx.doi.org/10.1016/S0550-3213(97)00125-9}{{\em Nucl.Phys.}
  {\bfseries B493} (1997) 101--147},
\href{http://arxiv.org/abs/hep-th/9611063}{{\ttfamily arXiv:hep-th/9611063
  [hep-th]}}.

\bibitem{Porrati:1996xi}
M.~Porrati and A.~Zaffaroni, ``{M theory origin of mirror symmetry in
  three-dimensional gauge theories},''
  \href{http://dx.doi.org/10.1016/S0550-3213(97)00061-8}{{\em Nucl.Phys.}
  {\bfseries B490} (1997) 107--120},
\href{http://arxiv.org/abs/hep-th/9611201}{{\ttfamily arXiv:hep-th/9611201
  [hep-th]}}.

\bibitem{deBoer:1996ck}
J.~de~Boer, K.~Hori, H.~Ooguri, Y.~Oz, and Z.~Yin, ``{Mirror symmetry in
  three-dimensional theories, SL(2,Z) and D-brane moduli spaces},''
  \href{http://dx.doi.org/10.1016/S0550-3213(97)00115-6}{{\em Nucl.Phys.}
  {\bfseries B493} (1997) 148--176},
\href{http://arxiv.org/abs/hep-th/9612131}{{\ttfamily arXiv:hep-th/9612131
  [hep-th]}}.

\bibitem{Kapustin:1998fa}
A.~Kapustin, ``{D(n) quivers from branes},''
  \href{http://dx.doi.org/10.1088/1126-6708/1998/12/015}{{\em JHEP} {\bfseries
  9812} (1998) 015},
\href{http://arxiv.org/abs/hep-th/9806238}{{\ttfamily arXiv:hep-th/9806238
  [hep-th]}}.

\bibitem{Hanany:1999sj}
A.~Hanany and A.~Zaffaroni, ``{Issues on orientifolds: On the brane
  construction of gauge theories with SO(2n) global symmetry},'' {\em JHEP}
  {\bfseries 9907} (1999) 009,
\href{http://arxiv.org/abs/hep-th/9903242}{{\ttfamily arXiv:hep-th/9903242
  [hep-th]}}.

\bibitem{Hanany:2001iy}
A.~Hanany and J.~Troost, ``{Orientifold planes, affine algebras and magnetic
  monopoles},'' {\em JHEP} {\bfseries 0108} (2001) 021,
\href{http://arxiv.org/abs/hep-th/0107153}{{\ttfamily arXiv:hep-th/0107153
  [hep-th]}}.

\bibitem{Julia:1982gx}
B.~Julia,
``{Kac-Moody symmetry of gravitation and supergravity theories},''.

\bibitem{Sen:1998ii}
A.~Sen, ``{Stable nonBPS bound states of BPS D-branes},''
  \href{http://dx.doi.org/10.1088/1126-6708/1998/08/010}{{\em JHEP} {\bfseries
  9808} (1998) 010},
\href{http://arxiv.org/abs/hep-th/9805019}{{\ttfamily arXiv:hep-th/9805019
  [hep-th]}}.

\bibitem{'tHooft:1977hy}
G.~'t~Hooft, ``{On the Phase Transition Towards Permanent Quark Confinement},''
\href{http://dx.doi.org/10.1016/0550-3213(78)90153-0}{{\em Nucl.Phys.}
  {\bfseries B138} (1978) 1}.

\bibitem{Englert:1976ng}
F.~Englert and P.~Windey, ``{Quantization Condition for 't Hooft Monopoles in
  Compact Simple Lie Groups},''
\href{http://dx.doi.org/10.1103/PhysRevD.14.2728}{{\em Phys.Rev.} {\bfseries
  D14} (1976) 2728}.

\bibitem{Goddard:1976qe}
P.~Goddard, J.~Nuyts, and D.~I. Olive, ``{Gauge Theories and Magnetic
  Charge},''
\href{http://dx.doi.org/10.1016/0550-3213(77)90221-8}{{\em Nucl.Phys.}
  {\bfseries B125} (1977) 1}.

\bibitem{Kapustin:2005py}
A.~Kapustin, ``{Wilson-'t Hooft operators in four-dimensional gauge theories
  and S-duality},'' \href{http://dx.doi.org/10.1103/PhysRevD.74.025005}{{\em
  Phys.Rev.} {\bfseries D74} (2006) 025005},
\href{http://arxiv.org/abs/hep-th/0501015}{{\ttfamily arXiv:hep-th/0501015
  [hep-th]}}.

\bibitem{Borokhov:2002cg}
V.~Borokhov, A.~Kapustin, and X.-k. Wu, ``{Monopole operators and mirror
  symmetry in three-dimensions},'' {\em JHEP} {\bfseries 0212} (2002) 044,
\href{http://arxiv.org/abs/hep-th/0207074}{{\ttfamily arXiv:hep-th/0207074
  [hep-th]}}.

\bibitem{Borokhov:2003yu}
V.~Borokhov, ``{Monopole operators in three-dimensional N=4 SYM and mirror
  symmetry},'' \href{http://dx.doi.org/10.1088/1126-6708/2004/03/008}{{\em
  JHEP} {\bfseries 0403} (2004) 008},
\href{http://arxiv.org/abs/hep-th/0310254}{{\ttfamily arXiv:hep-th/0310254
  [hep-th]}}.

\bibitem{Benna:2009xd}
M.~K. Benna, I.~R. Klebanov, and T.~Klose, ``{Charges of Monopole Operators in
  Chern-Simons Yang-Mills Theory},''
  \href{http://dx.doi.org/10.1007/JHEP01(2010)110}{{\em JHEP} {\bfseries 1001}
  (2010) 110},
\href{http://arxiv.org/abs/0906.3008}{{\ttfamily arXiv:0906.3008 [hep-th]}}.

\bibitem{Bashkirov:2010kz}
D.~Bashkirov and A.~Kapustin, ``{Supersymmetry enhancement by monopole
  operators},'' \href{http://dx.doi.org/10.1007/JHEP05(2011)015}{{\em JHEP}
  {\bfseries 1105} (2011) 015},
\href{http://arxiv.org/abs/1007.4861}{{\ttfamily arXiv:1007.4861 [hep-th]}}.

\bibitem{kronheimer1990}
P.~B. Kronheimer, ``Instantons and the geometry of the nilpotent variety,''
  \href{http://projecteuclid.org/euclid.jdg/1214445316}{{\em Journal of
  Differential Geometry} {\bfseries 32} no.~2, (1990) 473--490}.

\bibitem{Br}
R.~Brylinski, ``{Instantons and K\"ahler geometry of nilpotent orbits},'' in
  {\em {Representation theories and algebraic geometry}}, vol.~514 of {\em NATO
  Adv. Sci. Inst. Ser. C Math. Phys. Sci.}, pp.~85--125.
\newblock Kluwer, 1998.
\newblock \href{http://arxiv.org/abs/math.SG/9811032}{{\ttfamily
  math.SG/9811032}}.

\bibitem{KS1}
P.~Kobak and A.~Swann, ``{The hyperk\"ahler geometry associated to Wolf
  spaces},'' {\em Boll. Unione Mat. Ital. Serie 8, Sez. B Artic. Ric. Mat.}
  {\bfseries 4} (2001) 587,
  \href{http://arxiv.org/abs/math.DG/0001025}{{\ttfamily math.DG/0001025}}.

\bibitem{VinbergPopov}
E.~B. Vinberg and V.~L. Popov, ``{On a class of quasihomogeneous affine
  varieties},'' \href{http://dx.doi.org/10.1070/IM1972v006n04ABEH001898}{{\em
  Math. USSR-Izv.} {\bfseries 6} (1972) 743}.

\bibitem{Garfinkle}
D.~Garfinkle, ``{A new construction of the Joseph ideal},'' 1982.
\newblock \url{http://hdl.handle.net/1721.1/15620}.

\bibitem{joseph1976minimal}
A.~Joseph, ``The minimal orbit in a simple lie algebra and its associated
  maximal ideal,'' {\em Ann. Sci. {\'E}cole Norm. Sup.(4)} {\bfseries 9} no.~1,
  (1976) 1--29.

\bibitem{Gaiotto:2008nz}
D.~Gaiotto, A.~Neitzke, and Y.~Tachikawa, ``{Argyres-Seiberg duality and the
  Higgs branch},'' \href{http://dx.doi.org/10.1007/s00220-009-0938-6}{{\em
  Commun.Math.Phys.} {\bfseries 294} (2010) 389--410},
\href{http://arxiv.org/abs/0810.4541}{{\ttfamily arXiv:0810.4541 [hep-th]}}.

\bibitem{Bashkirov:2010hj}
D.~Bashkirov, ``{Examples of global symmetry enhancement by monopole
  operators},''
\href{http://arxiv.org/abs/1009.3477}{{\ttfamily arXiv:1009.3477 [hep-th]}}.

\bibitem{Kim:2009wb}
S.~Kim, ``{The Complete superconformal index for N=6 Chern-Simons theory},''
  \href{http://dx.doi.org/10.1016/j.nuclphysb.2012.07.015,
  10.1016/j.nuclphysb.2009.06.025}{{\em Nucl.Phys.} {\bfseries B821} (2009)
  241--284},
\href{http://arxiv.org/abs/0903.4172}{{\ttfamily arXiv:0903.4172 [hep-th]}}.

\bibitem{Imamura:2011su}
Y.~Imamura and S.~Yokoyama, ``{Index for three dimensional superconformal field
  theories with general R-charge assignments},''
  \href{http://dx.doi.org/10.1007/JHEP04(2011)007}{{\em JHEP} {\bfseries 1104}
  (2011) 007},
\href{http://arxiv.org/abs/1101.0557}{{\ttfamily arXiv:1101.0557 [hep-th]}}.

\bibitem{Kapustin:2011jm}
A.~Kapustin and B.~Willett, ``{Generalized Superconformal Index for Three
  Dimensional Field Theories},''
\href{http://arxiv.org/abs/1106.2484}{{\ttfamily arXiv:1106.2484 [hep-th]}}.

\bibitem{Dey:2014tka}
A.~Dey, A.~Hanany, P.~Koroteev, and N.~Mekareeya, ``{Mirror Symmetry in Three
  Dimensions via Gauged Linear Quivers},''
\href{http://arxiv.org/abs/1402.0016}{{\ttfamily arXiv:1402.0016 [hep-th]}}.

\end{thebibliography}\endgroup

\end{document}